 \documentclass[referee]{aa}  
\usepackage{graphicx}
\usepackage{txfonts}
\begin{document} 

   \title{Phase Curves of Small Bodies from the SLOAN Moving Objects Catalog}


   \author{A. Alvarez-Candal
          \inst{1,2,3}
          \and
          P. G. Benavidez
          \inst{1,4}
          \and
          A. Campo Bagatin
          \inst{1,4}
          \and
          T. Santana-Ros
          \inst{4,5}
          }

   \institute{Instituto de F\'isica Aplicada a las Ciencias y las Tecnolog\'ias, Universidad de Alicante, San Vicent del Raspeig, E03080, Alicante, Spain\\
              \email{varobes@gmail.com}
        \and
            Instituto de Astrof\'isica de Andaluc\'ia, CSIC, Apt 3004, E18080 Granada, Spain
         \and
             Observat\'orio Nacional / MCTIC, Rua General Jos\'e Cristino 77, Rio de Janeiro, RJ, 20921-400, Brazil
        \and
            Departamento de F\'isica Ingenier\'ia de Sistemas y Teor\'ia de la Se\~nal, Universidad de Alicante, San Vicent del Raspeig, E03080, Alicante, Spain
        \and
            Institut de Ci\`encies del Cosmos (ICCUB), Universitat de Barcelona (IEEC-UB), Mart\'i Franqu\`es 1, E08028, Barcelona, Spain
             }

   \date{Received XX; accepted XX}

 
  \abstract
  {Extensive photometric surveys are and will continue producing massive amounts of data on small bodies. Usually, these data will be sparsely obtained at arbitrary (and unknown)rotational phases. Therefore, new methods to process such data need to be developed to make the most of those large catalogs.}
  {We aim to produce a method to create phase curves of small bodies considering the uncertainties introduced by the nominal errors in the magnitudes and the effect introduced by rotational variations.
  
  We use the SLOAN Moving Objects Catalog data as a benchmark to construct phase curves of all small bodies in there, in u', g', r', i', and z' filters. We will obtain from the phase curves the absolute magnitudes and set up with them the absolute colors, which are the colors of the asteroids not affected by changes in phase angle.}
  {We select objects with $\geq3$ observations taken in at least one filter and spanned over a minimum of 5 degrees in phase angle. We developed a method that combines Monte Carlo simulations and Bayesian inference to estimate the absolute magnitudes using the HG$_{12}^*$ photometric system.}
  {We obtained almost 15\,000 phase curves, about 12\,000 including all five filters. The absolute magnitudes and absolute colors are compatible with previously published data, supporting our method.}
  {The method we developed is fully automatic and well suited to be run on large amounts of data. Moreover, it includes the nominal uncertainties in the magnitudes and the whole distribution of possible rotational states of the objects producing, possibly, less precise values, i.e., larger uncertainties, but more accurate, i.e., closer to the actual value. To our knowledge, this work is the first to include the effect of rotational variations in such a manner.}

   \keywords{Methods: data analysis -- Catalogs -- Minor planets, asteroids: general
}
   \titlerunning{Multi-wavelength phase curves of asteroids}
   \authorrunning{Alvarez-Candal et al.}

   \maketitle
%

\section{Introduction}\label{sec:intro}

The field of modern planetary sciences is experiencing a revolution due to: (1) Space missions to targets of interest and (2) massive databases produced as outcomes from extensive observational surveys. As far as item (1) is concerned, we will only mention that, although extremely useful, space missions provide detailed information about a handful of objects. However, these data are used to impose constraints on whole populations [for instance 67P, visited by ESA's Rosetta mission \cite[][and references therein]{barucci2017rosseta}\footnote{https://sci.esa.int/web/rosetta/}, or JAXA's Hayabusa visit to Itokawa \cite[][and references therein]{fujiwara2006Sci}\footnote{https://global.jaxa.jp/projects/sas/muses\_c/} it is like admiring a painting at a very close range. On the other hand, extensive observational surveys produce large amounts of data, which allow us to study whole populations, their distribution in the observational parameters space, or relations, while being more difficult to obtain detailed information about individual objects, usually because most of them would only be observed once or twice. This is like admiring a painting from a distance, recognizing its patterns but not the details of the strokes. We present a method to make the most of data provided by large observational surveys in this work.

Among the future and ongoing surveys, some of them will produce multi-wavelength photometric catalogs.
Two examples are, first, the future Legacy Survey of Space and Time (LSST) to be executed by the Vera C. Rubin Observatory, which will observe massive numbers of small bodies \citep{jones2009lsst}, and the already published MOVIS catalog that gathers near-infrared magnitudes of small bodies from the VISTA-VHS survey \citep{popescu2016movis}. In particular, for the scope of the present work, we are interested in two specific surveys: The Javalambre Physics of the Accelerating Universe, J-PAS \citep{benitez2014arXiv}\footnote{http://www.j-pas.org} that will observe 8\,000 sq degrees of the northern sky in 56 different filters (54 narrowband and two broadband) using a 2.5-m telescope and the Javalambre Photometric Local Universe Survey, J-PLUS \citep{cenarro2019}\footnote{http://www.j-plus.es} that is observing the same footprint as J-PAS but with 12 filters (5 broadband) using a 0.8-m telescope. The power of these surveys relies not only on the large footprint but also on the number of filters to be used for the observations, which provides unprecedented spectral coverage for spectro-photometry \citep[see][for an example on J-PLUS data of asteroids]{Morate2021arXiv}. Unfortunately, the data are taken sparsely due to the observational strategy, and different filters may be observed at different epochs. Therefore, new methods need to be developed to make these data useful for small bodies.

Aiming at providing one such tool, we develop a method that combines Monte Carlo simulations and Bayesian inference to include the effect of rotational variation between different epochs. As a test, and also because of its importance in the study of small bodies, we concentrate on data produced by the SLOAN Digital Survey \citep{york2000AJ}, whose photometric survey produced detection of several thousand moving objects, many of them linked to real objects in their Moving Objects Catalog \cite[hereafter just MOC,][]{ivezic2001AJ,juric2002AJ}. The MOC data have been widely used to characterize diverse populations of asteroids, comets, and trans-Neptunian objects (TNOs). Some applications have been the taxonomical classification of small bodies and the study of the mass distribution in the Main Belt of asteroids \citep[for example:][and others]{Roig2006Icar,Roig2008A&A,gh2008Icar,assandri2008,gh2010Icar,carvano2010,solontoi2010Icar,ofek2012ApJ,solontoi2012Icar,alcan2013,demeo2013Icar,demeo2014Natur,hasselmann2015,carry2016Icar}. In particular, we concentrate on those objects that appear multiple times in the MOC ($>15\,000$ with at least three different observations) in any of the five filters of the SLOAN photometric system. We aim at constructing their {\it phase curves} in all filters used by the survey, providing a catalog of zero-phase magnitudes or absolute magnitudes. This work does not aim to perform an in-depth analysis of the results but to provide the community with a valuable set of data to be further studied.

The absolute magnitude, $H$, of an airless small body is defined as the object's apparent magnitude if illuminated by the Sun at 1 AU while being observed from a distance of 1 AU at opposition. The importance of $H$ comes from its well-known relation with the size of the object and its albedo \citep{bowelllumme1979}. $H$ is computed using {phase curves (PCs hereafter)} which relate the change of the apparent magnitudes with the phase angle ($\alpha$, which is the angular distance between the Sun and the Earth as seen from the object). Due to the characteristics of the PCs, it is necessary to (i) have observations at several different $\alpha$, and (ii) to take into account the effect of the rotational variability. Therefore, a PC usually cannot be constructed with data of only one observational run (spanning just a few nights), except for Near Earth Objects (NEOs), because there may not be enough coverage in $\alpha$. Note that possible changes in brightness due to different apparitions (aspect angle-related) should also be included. Nevertheless, this issue will not be tackled in this work.

Two different approaches can be used to create PCs. On the one hand, targeted observations of selected objects, with dedicated telescope (or telescopes) time \cite[][among others]{carbo2019}. This approach has obvious advantages: target selection, enough time to remove rotational variation (we will come back to this in Sect. \ref{sec:analysis}), and control over the overall observational setup (i.e., filter selection, range of $\alpha$, etc.) On the downside, these observations can demand many telescope hours and careful planning to have maximum efficiency. On the other hand, it is possible to use the large amount of data produced by several extensive observational surveys (on-going and future). Key references in this regard are the almost half a million absolute magnitudes from the Minor Planets Center compiled by \cite{oszki2011pcs}, the 250\,000 Pan-STARRS absolute magnitudes \citep{veres2015},  the analysis of almost 100\,000 asteroids from the ATLAS survey \citep{mahlke2021}, or the almost 15\,000 PCs from Gaia Data Release 2 (DR2) observations \citep{colazo2021}. This second approach has no control over the observational setup or target selection, and all work should be executed on serendipitous observations.  The advantage relies on massive numbers, several orders of magnitude larger than anything that can be done with targeted studies.

We aim in this work at constructing all possible PCs from the MOC for objects with at least three observations (different $\alpha$, and with a minimum spread of 5 degrees) in any of the filters. We obtain absolute magnitudes of several thousand objects that allow constructing colors at $\alpha=0$ degrees, i.e., not affected by phase coloring, which could serve as templates to study these effects. Therefore, in this work, we will use data from MOC (described in Sect. \ref{sect:data}) to create multi-wavelength phase curves (to the best of our knowledge, the only other similar approach was performed by \citealt{mahlke2021} using the cyan and orange filters of ATLAS). In order to do so, we have devised a method to account for rotational variation using concepts of Bayesian statistics (Sect. \ref{sec:analysis}). The results are presented and discussed in Sect. \ref{sec:discussion}. Finally, we draw our conclusions in Sect. \ref{sect:conclussions}.

\section{Dataset}\label{sect:data}
We use the $4^{th}$ release of the MOC\footnote{http://faculty.washington.edu/ivezic/sdssmoc/sdssmoc.html} that contains 471\,569 moving objects, 220\,101 linked to 104\,449 small bodies. This release (the last one of the MOC, dated from 2007) includes data obtained under non-photometric conditions. We increased this database with the updated SVOMOC version \citep{carry2016Icar} adding 57\,646 new observations linked to 36\,939 previously unidentified objects\footnote{Care must be taken when using this database because we identified 8\,617 repeated lines.}. For simplicity, we will keep the name of MOC for the input database. 

The MOC contains the magnitudes $u'$, $g'$, $r'$, $i'$, and $z'$\footnote{Note that we drop the $'$ in the remaining of the text.} and their respective errors, $\sigma_m$ (in the remaining of the text, we will use $m$ to name the apparent magnitude in any of the five SLOAN filters or the V filter unless explicitly stated. We will use italics when speaking about magnitudes and regular text when naming a filter). The catalog also provides the heliocentric $(R)$ and the topocentric $(\Delta)$ distances and $\alpha$ at the time of observation. We do not apply any cut in the database, except the removal of any $m : \sigma_{m}\geq1 $. We used only this quality-cut because the method is intended to be run automatically and provide, even in cases with poor coverage of the PCs or poor quality data, with a useful distribution of solutions that can later be used as priors once the databases are updated. To analyze the dataset, we devised a Python algorithm, based on our earlier work \cite[e.g.,][]{alcan2016,alcan2019}, that selects objects with at least three different $\alpha$ and with a minimum span of $\Delta\alpha\geq5.0$ degrees (i.e., the distance between the maximum and the minimum $\alpha$). We chose this value using a trial-and-error approach as a good balance between the number of objects and the quality of the results. In order to assess the reliability of our results, we will compare them with four different sets of $H$, the first one from data compiled from the Minor Planets Center \citep{oszki2011pcs}, the second one from Pan-STARRS \citep{veres2015}, the third one from Gaia data \citep{colazo2021}, and the last one from the ATLAS survey \citep{mahlke2021}.

\section{Analysis}\label{sec:analysis}
To obtain the PCs, we first select a photometric model to fit the observational data. Among the several available in the literature (for example, see \citealt{carbo2019} for a comparison of three of them) we will use the HG$_{12}^*$ model as described in \cite{penti2016HG} because most objects have a sparse coverage of the PCs. The HG$_{12}^*$ model is adapted from the HG$_1$G$_2$ \citep{muinonen2010HG1G2} and should not be confused with the HG$_{12}$.

We stress that the observations in the MOC correspond to serendipitous observations of small bodies while they were crossing the field of view of the survey. Several thousand objects were observed more than once at different $\alpha$, which permitted us to construct PCs in the filter system. Although observations in one run (one set of ${m}$) are taken almost simultaneously, and rotational variations could be disregarded, this is not the case for observations at different $\alpha$, that were taken at an arbitrary rotational phase of the object. This is a crucial effect to be considered when creating PCs by joining data of different epochs (among other possible effects that will be mentioned below and in Sect. \ref{sect:caveats}). In the following subsections, we discuss the computation of the PCs first and how we include the effect of the unknown amplitude of the rotational light curve ($\Delta m$).

\subsection{Computation of the phase curve}
The adopted form of the phase curves is 
\begin{equation}\label{eq:pcurve}
M(\alpha)=H-2.5\log{[G_1\Phi_1(\alpha)+G_2\Phi_2(\alpha)+(1-G_1-G_2)\Phi_3(\alpha)]},
\end{equation} 
where $M(\alpha)=m-5\log{(R\Delta)}$ is the reduced magnitude, $H$ is the absolute magnitude, $\Phi_i$ are known functions of $\alpha$ (see \citealt{penti2016HG} for the tables), and $G_i$ are the phase coefficients that provide the shape of the curve. In the HG$_{12}^*$ model, $G_1$ and $G_2$ are parametrized by $G_{12}^*$ according to $G_1=0.84293649\times G_{12}^*$ and $G_2=0.53513350\times(1-G_{12}^*)$. From now on, we will call $G_{12}^*$ as $G$ for simplicity unless explicitly stated otherwise, and we will not use sub-indexes to differentiate among filters unless necessary. We use the script written in Python made public by \cite{penti2016HG}\footnote{https://wiki.helsinki.fi/display/PSR/HG1G2+tools} to compute the parameters $H$ and $G$, given the values of $M(\alpha)$ and corresponding errors. The code returns the nominal values and an estimate of their errors based on a Monte Carlo simulation. However, we will not use their error estimates but instead compute them from our own Monte Carlo simulations because ours will consider the nominal errors of the magnitudes and the possible rotational state of the objects (as described in the following Sects.). 

\subsubsection{First test}
\begin{figure}
\centering
\includegraphics[width=\hsize]{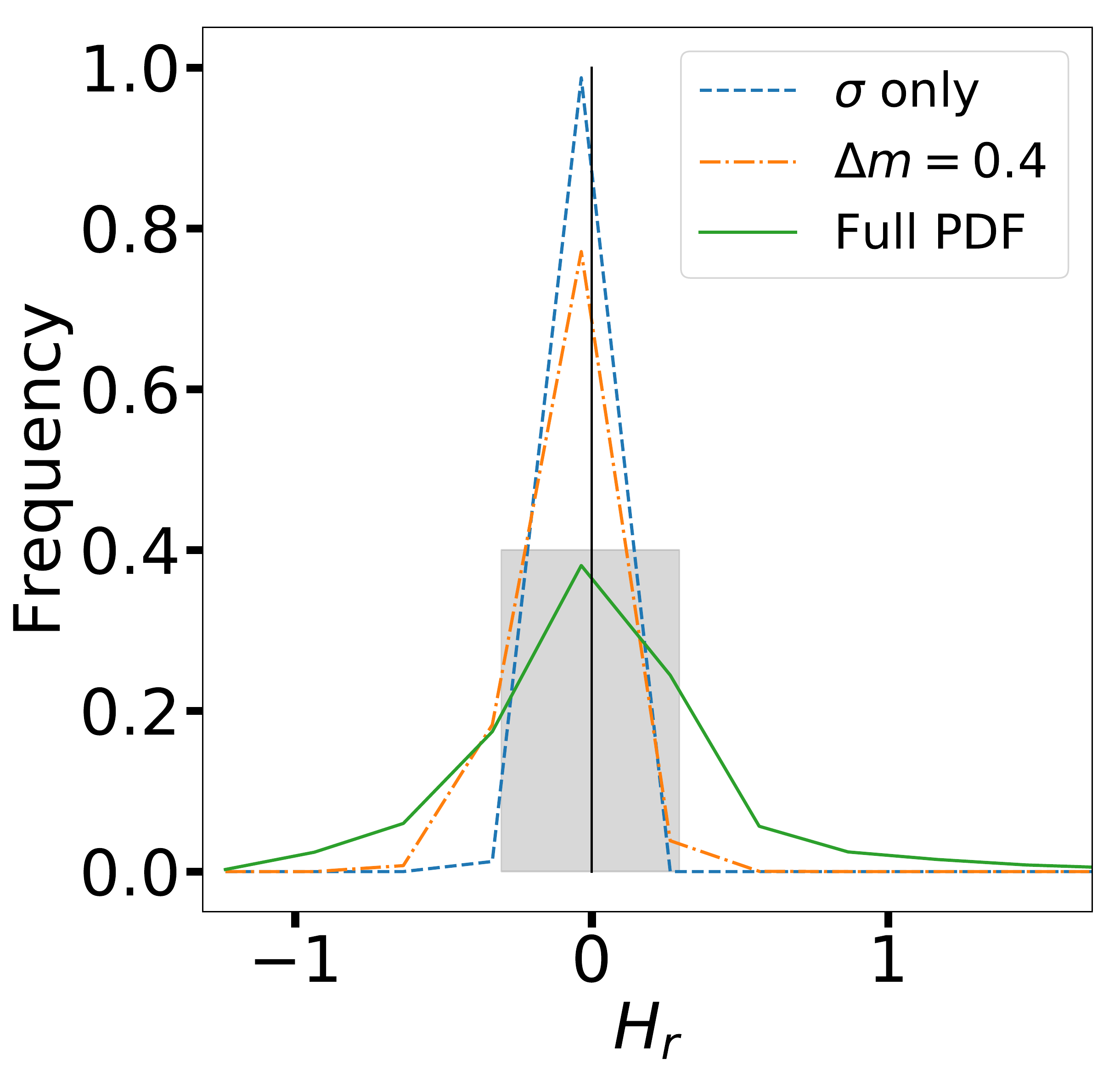}
\caption{Comparison between three different Probability Distribution Functions for the $H_r$ of 1034 T-1. All distributions are centered at the median value of the magnitude obtained with the Full PDF. The blue dashed line considers only the errors in the magnitude, the orange dot-dashed line considers a single $\Delta m=0.4$, and the  continuous green line considers all possible values of $\Delta m$ and the photometric errors (see text). The black vertical line represents the zero (i.e., the median of the Full PDF), while the shaded area covers the interval where the 68\% of the solutions lie in the green curve.}\label{fig:fig1}%
\end{figure}

We computed $H$ and $G$ for each object and filter using Eq. \ref{eq:pcurve}. We generated 10\,000 different solutions of the PCs for each run by varying each magnitude within their errors following $m*=m+rand_i$, where $rand_i$ is a random number extracted from a normal distribution with $\sigma=\sigma_{m}$. This is based on the method developed and tested in \cite{alcan2016,alcan2019} and \cite{ayala2018} for TNOs. Note that it may happen that some of the 10\,000 runs resulted in non-valid solutions. This means that the obtained $H$ and $G$ values result in PCs with no physical meaning, although being mathematically valid, for example, objects with abrupt drops (more than a couple of magnitudes) in brightness close to opposition, or objects with $G$ outside its interval of formal validity $[0,1]$ \citep[see][for details]{penti2016HG}. Therefore the final number of solutions is $N\leq10\,000$. The outcome of a run for our test subject, 1034 T-1\footnote{We have chosen 1034 T-1 as our test object because it was the first object that fulfilled our criteria in the original MOC catalog, without including SVOMOC data.}, in the r filter can be seen in Fig. \ref{fig:fig1} in the blue dashed line, labeled as {\it $\sigma$ only}.
In this case, $H_r$ is very precisely estimated (see Table \ref{table:1}). Nevertheless, the quoted uncertainty ($\sigma_{H_r}=0.04$ mag) does not reflect any possible effect of rotational variations in the magnitude, which will probably be larger than 0.04 based on its approximated diameter (between 3 and 6 km depending on its albedo), unless all observations were performed when the object was on a pole-on configuration. Nevertheless, we do not know this {\it a priori}, and it is unlikely that all observations were carried with such configuration.

\subsubsection{Second test}
The previous test did not include the effect of any possible rotational variation between different observations. This effect can be quite dramatic. According to the light-curve database \citep{wagner2019} there are objects with $\Delta m = 3.0$ mag, which may radically change the PCs if the effect is not taken into account. 
To test its effects, we ran a second test similar to the first, but in this case, instead of selecting $rand_i$ from a normal distribution, we selected it from a uniform distribution $\in[-\Delta m_{med}/2,\Delta m_{med}/2]$, where $\Delta m_{med}=0.4$ is the median $\Delta m$ from the whole small body population. The outcome of this test is also displayed in Fig. \ref{fig:fig1} but in orange dot-dashed line, labeled as {\it $\Delta m=0.4$}. In this case, the distribution of $H_r$ (called posterior distribution) is broader than in the previous test, $\sigma_{H_r}=0.14$ mag, which may appear as more realistic, but in this case, we are only considering one possible rotational state of the object, which may not be the correct one.

\subsubsection{Including $\Delta m$ and $\sigma_m$}\label{sec:full_pdf}
From the two previous tests, it is possible to see that including only the errors in the magnitudes ($\sigma_m$) provides a very precise value of $H$ (see also Table \ref{table:1}), but it does not represent the real spread of the data (please remember that we are dealing with sparse data taken at unknown rotational phases). If we only include the effect of $\Delta m$, the spread of the posterior distribution $H$ increases, providing less precision but possibly more accuracy, without significantly changing the mean (nor the median) values of $H$ (Table \ref{table:1}). Nevertheless, even if we chose a similar method as in \cite{alcan2019} where $\Delta m$ was chosen according to an estimate of $H$, this does not reflect the whole space of possible values that $\Delta m$ could take for an object with a given $H$. This is because different objects with similar $H$ may have different values of $\Delta m$. Therefore, we have adopted a new approach that includes the effect of the individual uncertainties in the magnitudes and also the whole space of possible values of $\Delta m$, i.e., it includes the values of $\Delta m$ for all objects with $H_V$ similar to the object of interest.
\begin{table}
\caption{Absolute magnitude estimates for asteroid 1034 T-1. The first column indicates the method, the second column the mean value, the third column shows the median value, and the fourth the standard deviation as a proxy for the width of the distributions}.
\label{table:1}[h]
\centering
\begin{tabular}{c c c c}
\hline\hline
Method              & $\overline{H_r}$ & $\tilde{H_r}$ & $\sigma_{H_r}$ \\
\hline
$\sigma$-only       & 14.95           & 14.92         & 0.04        \\
$\Delta m=0.4$      & 14.94           & 14.88         & 0.14        \\
Full PDF            & 14.99           & 14.99         & 0.45        \\
\hline
\end{tabular}
\end{table}

\begin{figure}
\centering
 \includegraphics[width=6.5cm]{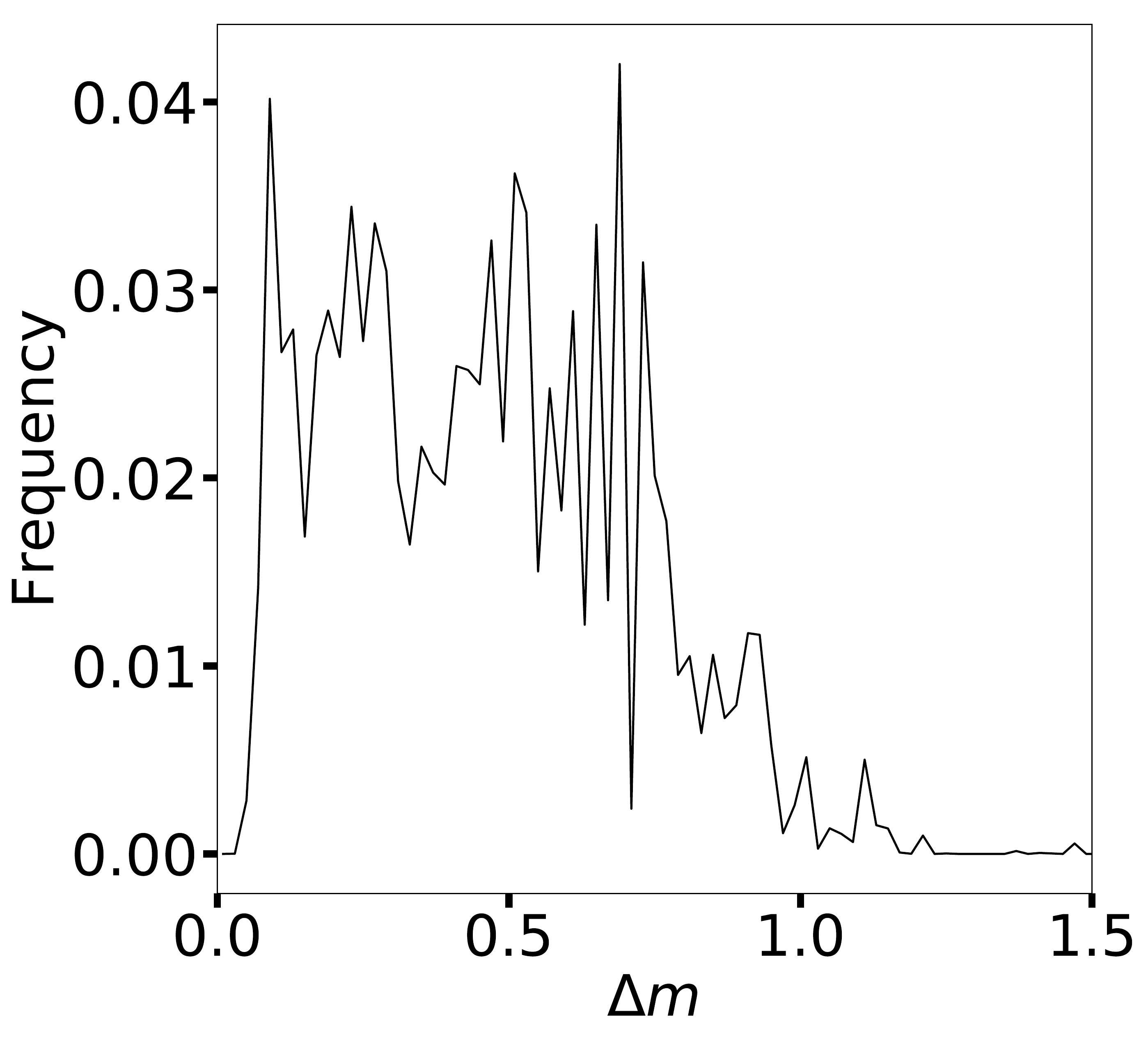}
 \includegraphics[width=6.5cm]{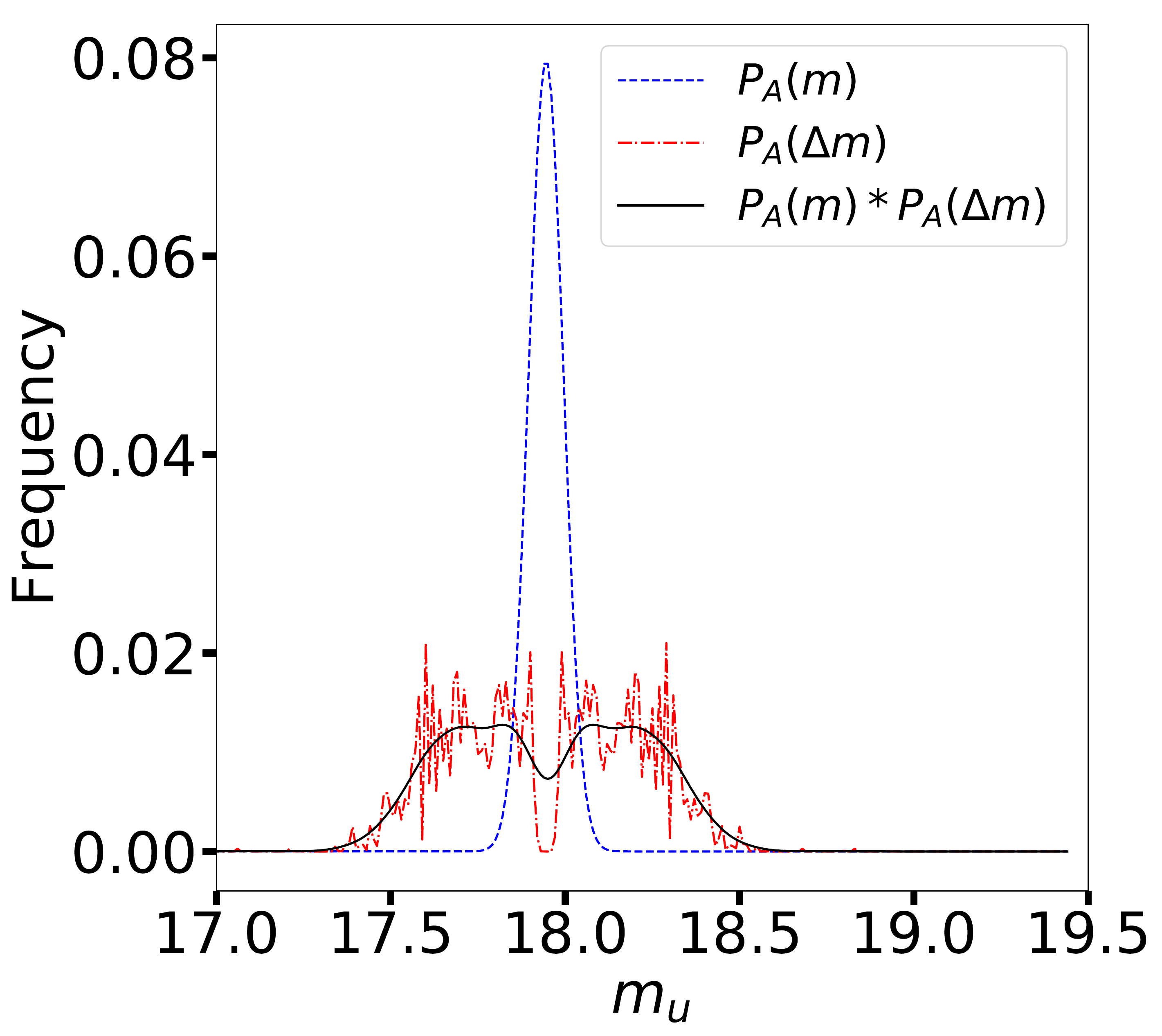}

\caption{Left: Example of the Probability Distribution Function $P_A(\Delta m)$, cut at 1.5 in the x-axis for clarity, for asteroid 1034 T-1. Right: Examples of the PDFs used in this work. }\label{fig:fig2}%
\end{figure}

\noindent
The method we devised includes the following steps (per object and per filter):\\
\noindent
$1^{st}$ -- using Eq. \ref{eq:pcurve} we ran 10\,000 different solutions following Test 1, i.e., using only $\sigma_m$, and applying it only to $g$ and $r$, therefore, we computed $H_{g}$ and $H_{r}$. Using the  transformations in Table 1 of \cite{jester2005AJ} for all stars with $R - I < 1.15$\footnote{See also http://www.sdss3.org/dr8/algorithms/sdssUBVRITransform.php} we computed $H_{V}$, obtaining its posterior distribution.  We normalized this distribution to obtain a probability distribution function (PDF) in 30 bins, i.e., $P(H_{V})=\{P_j(H_{V})\}$, $j$ is an integer $\in[1,30]$.  \\
\noindent
$2^{nd}$ -- from the distribution of $\Delta m$ {\it vs} $H_V$ \citep{wagner2019} we extracted all values $\Delta m$ in each bin $j$ in $P(H_{V})$ and finally we summed them using the values of $P(H_{V})$ as weights:
\begin{equation}\label{eq:two}
P_A^i(\Delta m)=
        \frac{\sum_j{P^i(\Delta m|P_j(H_{V}))\times P_j(H_{V})}}
             {\sum_i{\sum_j{P^i(\Delta m|P_j(H_{V}))\times P_j(H_{V})}}}
             ,
\end{equation}
where $P_A(\Delta m)=\{P_A^i(\Delta m)\}$ is the PDF of $\Delta m$ for object $A$ in 150 bins of width 0.02 mag, to cover the whole range of reported $\Delta m\in[0.0,3.0]$. The denominator is a normalization factor. With this procedure, we created a PDF of possible values of $\Delta m$ that the object could have (see an example in Fig. \ref{fig:fig2}, left panel). \\
\noindent
$3^{rd}$ -- to create the final PDF that will be used to estimate the final range of possible solutions of $H$ in all filters, we must also include the uncertainty in the measurement of $m$. In this case we construct for each object and measurement the distribution $P_A(m)$ as a normal distribution with $\sigma=\sigma_m$ centered at $m$. The final PDF is the convolution of $P_A(m)$ and Eq. \ref{eq:two}:
\begin{equation}\label{eq:three}
    P_A(m,\Delta m)=\big(P_A(m)*P_A(\Delta m)\big).
\end{equation}
Note that, in order to include realistic values of $\Delta m$ in $P_A(m)$, the distribution in Fig \ref{fig:fig2} (left) was mirrored along the y-axis and scaled by a factor 0.5 in the x-axis. This is done because an object with a given $\Delta m$ takes values of $m$ in the domain $[m-\Delta m/2,m+\Delta m/2]$ (see the final distribution in Fig. \ref{fig:fig2}, right panel).
Extracting 10\,000 values randomly from $P_A(m,\Delta m)$ we ran the HG$^*_{12}$ model obtaining the final values of $H$. The obvious consequence of our procedure is that the distribution of $H$ is very wide (see Fig. \ref{fig:fig1} in the continuous green line, labeled as {\it Full PDF}, and also see the values in Table \ref{table:1}). We remark that this is not related to the precision of $H$ but its accuracy.

Our final results are $H$ and $G$ distributions for each object and filter (stored in a compressed ASCII file). The files are available upon request. An example for the $H_r$ for the asteroid 1034 T-1 is shown in Fig. \ref{fig:fig1}.
We chose our representative value\footnote{The value include a final cut $-0.1\leq G\leq1.1$. The stored files include all values.} (see Table \ref{table:4}) as the median of the distribution and the uncertainties as to the values that enclose 68\% of the solutions (seen in a continuous line and shaded region in the same figure).
\begin{table*}
\caption{Sample of the catalog showing the ID in the first column, $H_g$ and its uncertainties in the second to fourth columns, the number of observations considered in the fifth and $G_g$ and uncertainties from column six to eight. The last two columns show the minimum $\alpha$ and its total span. The complete catalog is available in the folder provided in the text. The information is split into three different files, one with the $H$ and $G$, one with the IDs, and the last one with the angles. This was done so for simplicity. Flag -9999 indicates objects with less than 3 observations in a given filter, while flag -8888 indicates that not enough valid solutions were obtained (at least 100). {\bf The complete catalog is available at the CDS (https://cdsarc.u-strasbg.fr/) or upon request.}}
\label{table:4}
\centering
\begin{tabular}{c| c c c c |c c c | c c }
\hline\hline
ID & $H_g$ & $\sigma_{H_g}^-$ & $\sigma_{H_g}^+$ & N & $G_g$ & $\sigma_{G_g}^-$ & $\sigma_{G_g}^+$ & $\alpha_{min}$ (deg) & $\Delta\alpha$ (deg) \\
\hline
1013 T-2 & 16.81   &  0.32&  0.41& 4& 0.56   & 0.40    & 0.35  &  7.42& 16.19\\
1034 T-1 & 15.75   &  0.22&  0.27& 4& 0.44   & 0.36    & 0.31  &  1.63& 22.44\\
1041 T-3 & 16.30.  &  0.38&  0.44& 3& 0.60   & 0.45    & 0.34  &  0.12&  8.36\\
1093 T-2 & -9999   & -9999& -9999& 2& -9999  & -9999   & -9999 &  4.13& 23.24\\
1132 T-1 & -8888   & -8888& -8888& 8& -8888  & -8888   & -8888 &  9.17& 20.86\\
\hline
\end{tabular}
\end{table*}

\section{Results and discussion}\label{sec:discussion}
We computed $H$ in all filters available in the MOC (see numbers in Table \ref{table:2}) following the procedure outlined above. In addition, we computed $H_V$ again using $V$ obtained from $g$ and $r$ for each $\alpha$ where both magnitudes were available. Using $V$ and the corresponding $\alpha$, we processed the data the same way as all the rest. These computations have a multi-fold objective: First, to compare our procedure with previously published works. In particular with \citet[][O11 from now on]{oszki2011pcs}, \citet[V15]{veres2015}, and \citet[C21]{colazo2021}. Second, to check the validity of using the HG$^*_{12}$ model, which is developed to obtain absolute magnitudes in the V filter. Third, for future studies that will include albedos in order to estimate diameters\footnote{This will not be presented in this work because our objective is to present a catalog of $H$s.}. We also include a comparison with \citet[M21]{mahlke2021} as other work with multi-wavelength PCs.
\begin{table}
\caption{Number of absolute magnitudes obtained}
\label{table:2}
\centering
\begin{tabular}{c c | c c}
\hline\hline
Magnitude  & N         & Concept & N \\
\hline
$H_u$   & 13\,299   & at least one valid $H  $   & 14\,801 \\
$H_g$   & 14\,081   & with the 5 $H$ valid       & 12\,252 \\
$H_r$   & 14\,543   & in common with O11         &  3\,879 \\
$H_i$   & 14\,485   & in common with V15         &  8\,050 \\
$H_z$   & 14\,104   & in common with C21         &     521 \\
$H_V$   & 14\,440   & in common with M21 (orange)&  2\,743 \\
        &           & in common with M21 (cyan)  &  1\,316 \\
\hline
\end{tabular}
\end{table}

\begin{figure}
\centering
\includegraphics[width=\hsize]{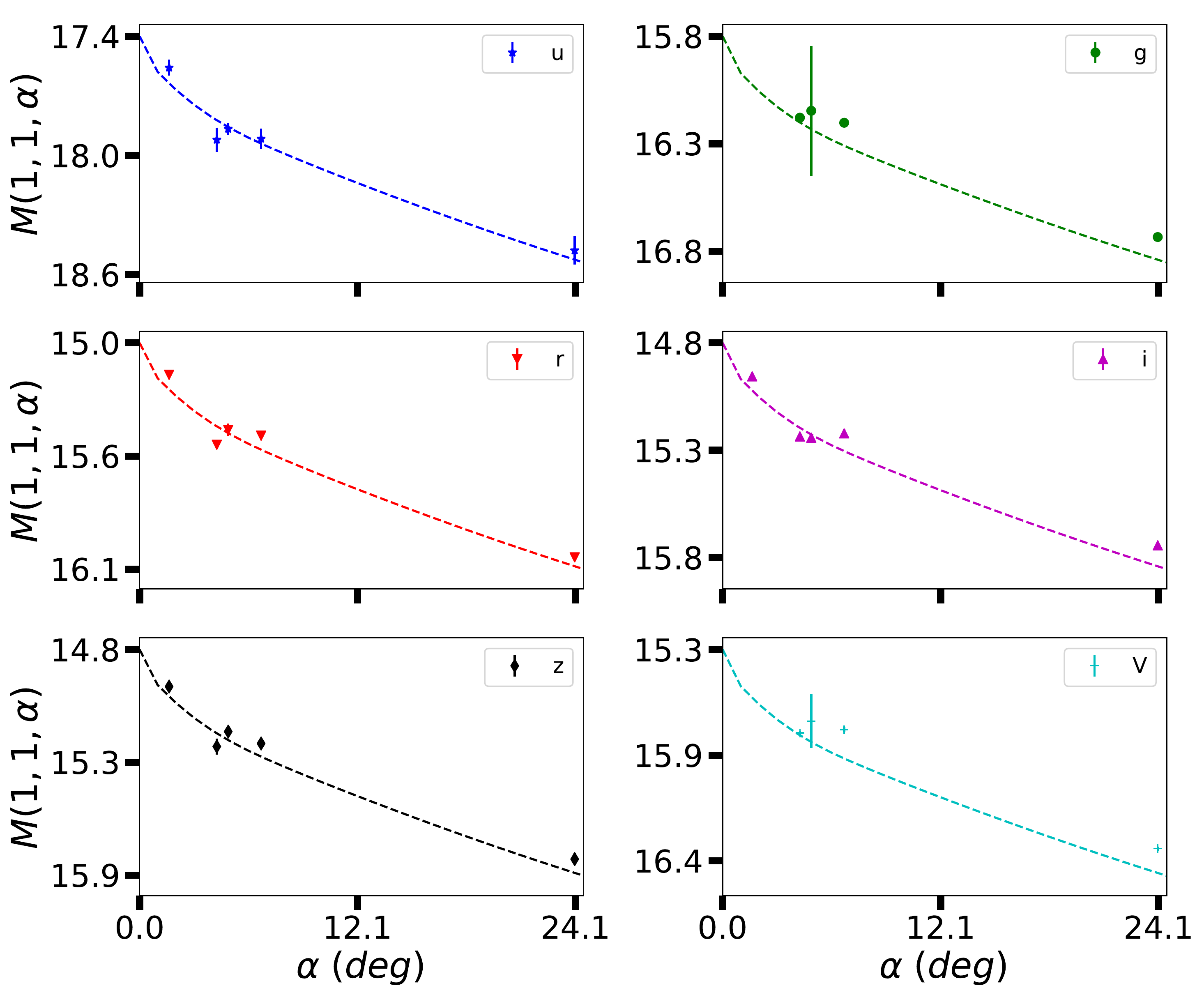}
\caption{Phase curves in all filters for asteroid 1034 T-1. The phase function correspond to the representative values. }\label{fig:fig10}%
\end{figure}

\begin{figure}
\centering
\includegraphics[width=\hsize]{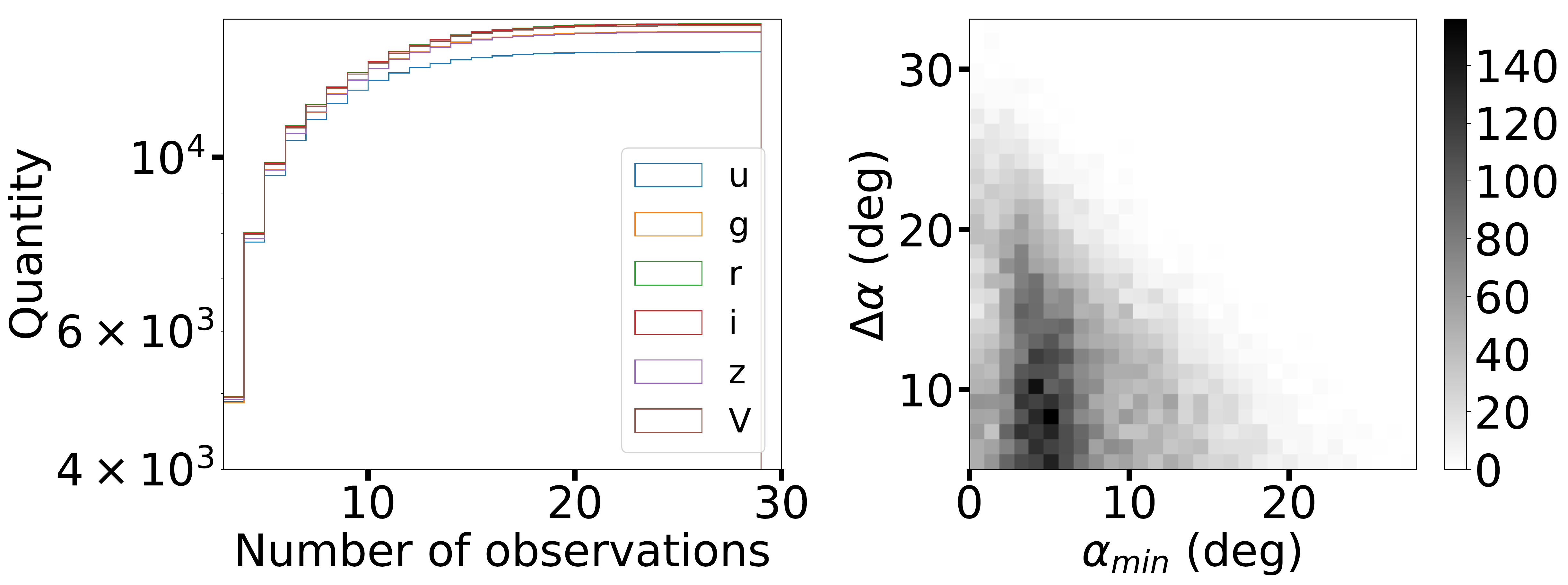}
\caption{Left panel: cumulative distribution of number observations per filter for the objects in this work. Right panel: minimum $\alpha$ vs. span in $\alpha$ covered by the objects in this work.}\label{fig:fig3}%
\end{figure}

\begin{figure}
\centering
\includegraphics[width=\hsize]{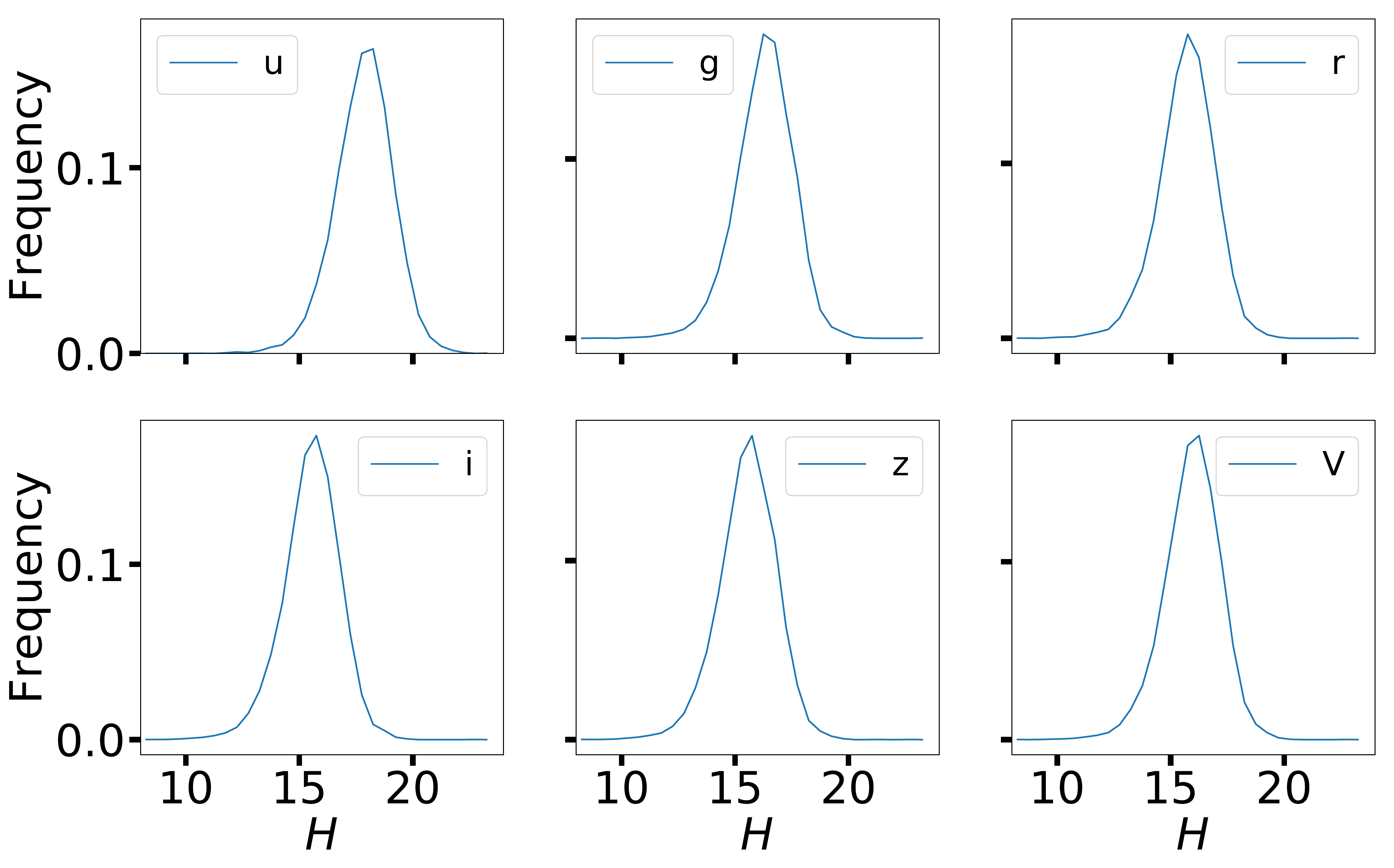}
\caption{Frequency distribution of $H_j$ obtained in this work.}\label{fig:fig9}%
\end{figure}

In Fig. \ref{fig:fig10} we show an example of the PCs for asteroid 1034 T-1 in the five filters of the SLOAN system and also in the transformed V filter. The curves represent the data satisfactorily, making us confident of the procedure we followed (see also Sect. \ref{sect:caveats} below).
From the sample point of view, most objects have only three observations, although there are some with more than ten as well (Fig. \ref{fig:fig3}, left panel), while the majority of objects are observed with $\alpha_{min}\sim5$ degrees. Note that there are objects observed from near opposition up to $\alpha\sim25$ degrees (Fig. \ref{fig:fig3}, right panel), being most of them Main Belt asteroids.
The distributions of absolute magnitudes in all filters are displayed in Fig. \ref{fig:fig9}.
From the figure, it is clear that the $H_u$ distribution is fainter than the others by about two magnitudes, which is compatible, but still slightly larger, with the solar $(u-g)=1.43$. Overall, a systematic decrease in brightness towards u filter indicates mostly positive spectral slopes in the $H_u-H_m$ colors. The solar colors could explain part of this systematic, but not all.

\subsection{HG$_1$G$_2$ vs. HG$^*_{12}$}\label{sec:models}
As described in \cite{penti2016HG}, the HG$^*_{12}$ model is well suited for PCs with sparse coverage, which is the case of our work, with some limitations for some taxonomical classes (see Sect. \ref{sect:caveats}). Nevertheless, recently, M21 showed that $H$ obtained from the HG$^*_{12}$ model are systematically offset from $H$ obtained using the full HG$_1$G$_2$ model. Their results were obtained using ATLAS data on about 100\,000 asteroids in two filters.

\begin{figure}
\centering
 \includegraphics[width=6.5cm]{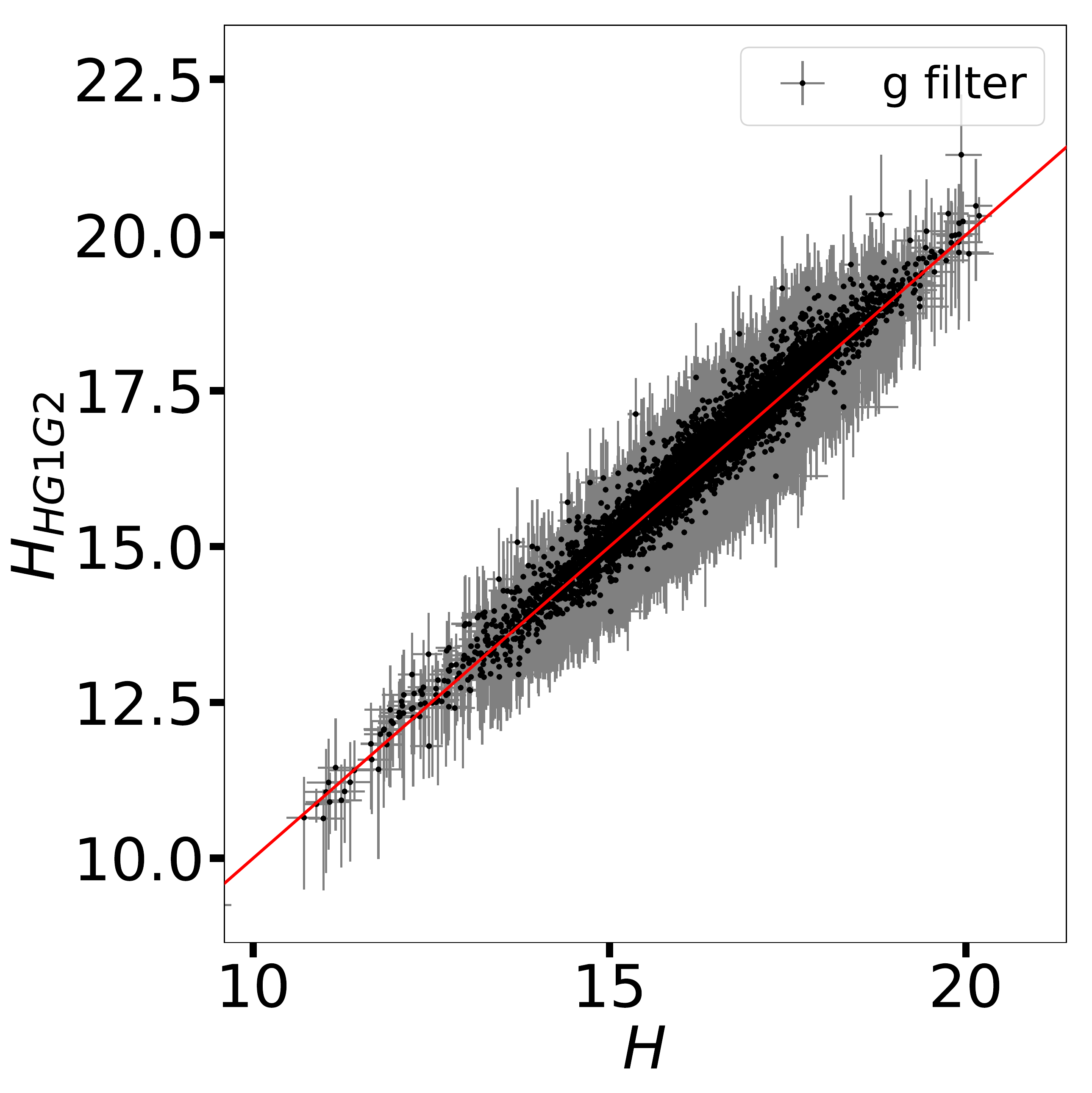}
 \includegraphics[width=6.5cm]{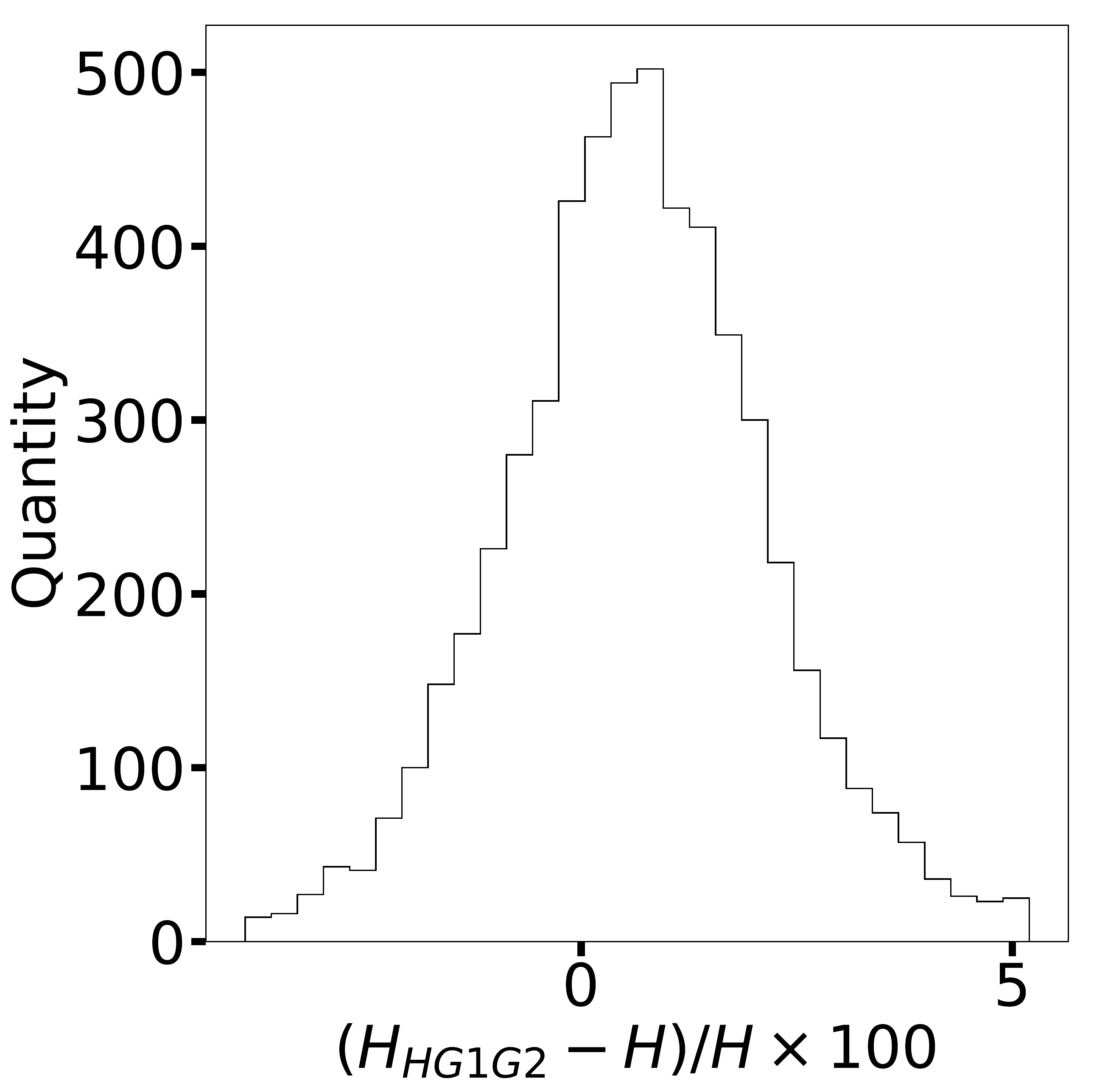}

\caption{Left panel: Comparison between $H_g$ 
obtained by us using the HG$_{12}^*$ (labeled $H$) and the magnitude obtained using the full HG$_1$G$_2$ models. The red line indicates the 1:1 relation. Right panel: Distribution of the percentual differences between the obtained $H$, the x-axis was cut for clarity.}\label{fig:Hgs}%
\end{figure}

To check whether our $Hs$ are trustworthy, we ran the whole code using the full HG$_1$G$_2$ model and compared the absolute magnitudes obtained. The results for $g$ are shown in Fig. \ref{fig:Hgs}; in all other filters, the results are qualitatively and quantitatively similar (see Table \ref{table:comparison}). We chose a percent representation of the data in the histograms to remove the impact of the actual value of $H$ in the comparisons, for instance, a 0.5 mag difference at $H=10$ implies a 5 \% difference, while at $H=20$ it implies a 2.5 \%. For an object with $H_g=16$, the median value in Table \ref{table:comparison} indicate a median difference of 0.05 magnitude, well below our median uncertainties. Therefore we conclude that $H$, as computed with the HG$_{12}^*$ model, is well suited for our purposes. This does not preclude that, in the future, and if the data better cover the PCs, we adopt the HG$_1$G$_2$ model.
\begin{table*}
\caption{Comparison of absolute magnitudes. The first column shows the magnitude, the second (and fourth) the number of objects in common, the third (and fifth) the median difference, the maximum, minimum and standard deviation.}
\label{table:comparison}
\centering
\begin{tabular}{c || c c | c c }
\hline\hline
Magnitude& N        & $(H_{HG1G2}-H)/H*100$ (\%)    & N        & $(H_{no\,OE}-H)/H*100$ (\%)    \\
         &          &  [median, min, $\sigma$, max] &          &  [median, min, $\sigma$, max] \\
\hline
$u$      & 5\,455   & -0.06, -23.7, 2.1,  9.7 &3\,242& 0.05, -32.7, 3.4, 12.7\\
$g$      & 7\,257   &  0.33, -24.7, 2.0, 11.5 &3\,515& 0.16, -37.4, 2.8, 12.9\\
$r$      & 7\,796   &  0.32, -26.5, 2.0, 10.8 &3\,644& 0.13, -40.9, 2.9, 11.6\\
$i$      & 7\,703   &  0.52, -28.0, 2.1, 11.3 &3\,615& 0.07, -41.2, 3.0, 12.2\\
$z$      & 6\,909   &  0.45, -27.8, 2.2, 11.8 &3\,526& 0.04, -40.8, 3.3,  9.2\\
$V$      & 7\,635   &  0.37, -25.8, 2.0, 12.6 &3\,611& 0.12, -40.7, 2.9,  9.0\\
\hline
\end{tabular}
\end{table*}

\begin{figure}
\centering
\includegraphics[width=\hsize]{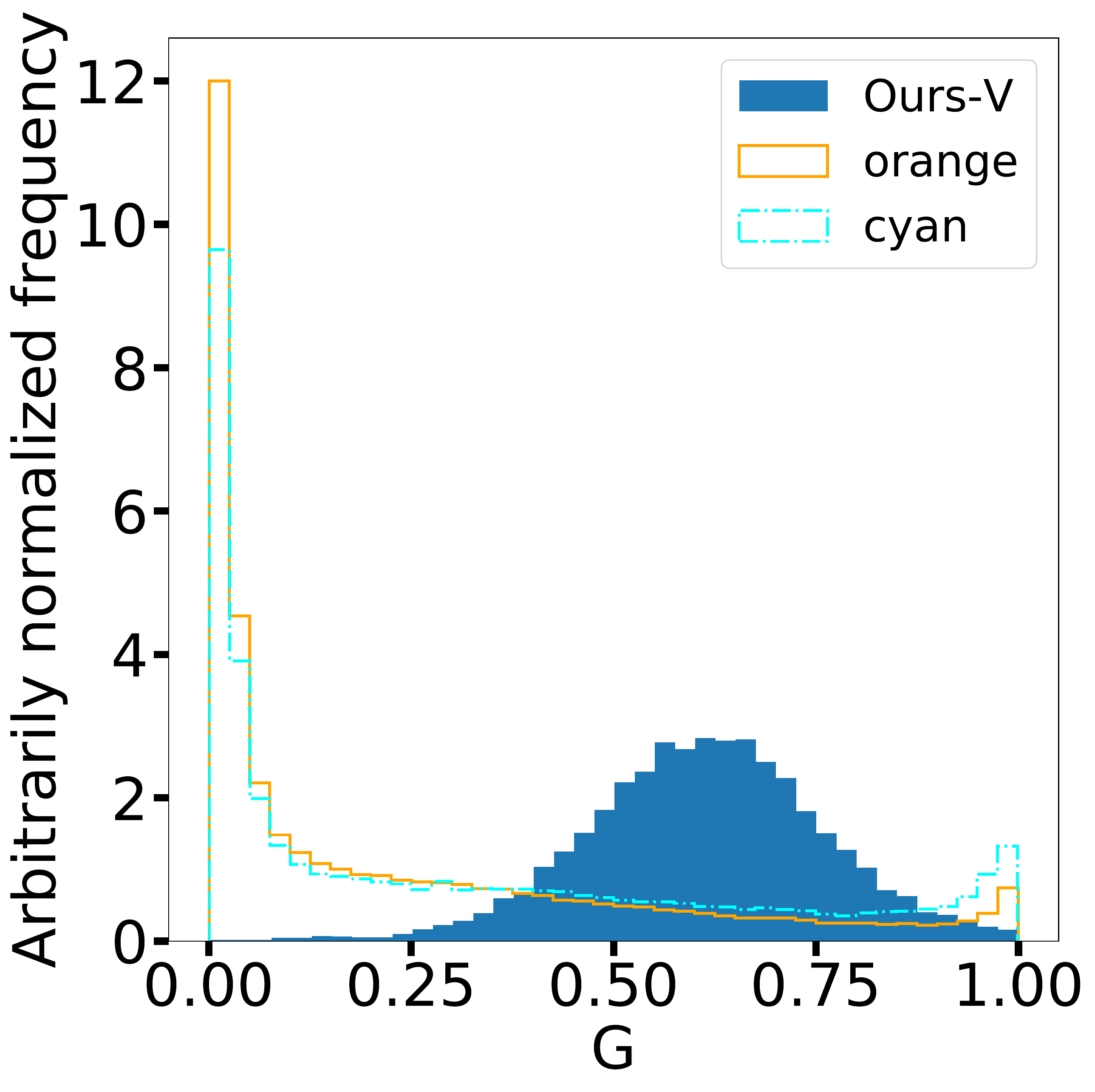}
\caption{Comparison between $G$ values, in different filters, obtained in M21 for the ATLAS survey (cyan and orange histograms) and $G_V$ in the blue histogram.}\label{fig:GAtlas}%
\end{figure}

M21 argued that the HG$_{12}^*$ was not precise enough because of a systematic difference in the absolute magnitudes obtained and also because the $G_{12}^*$ parameters they obtain tend to cluster towards 0 or 1 (see their Figs. 4 and 5). Nevertheless, we do not see the same trend. Our $G$-values have the shape of a wide bell. See, as an example, the comparison of $G_{cyan}$, $G_{orange}$, and our $G_V$ in Fig. \ref{fig:GAtlas}. 

\subsection{Opposition effect impact on our estimations of $H$}\label{sec:OE}

\begin{figure}
\centering
 \includegraphics[width=6.5cm]{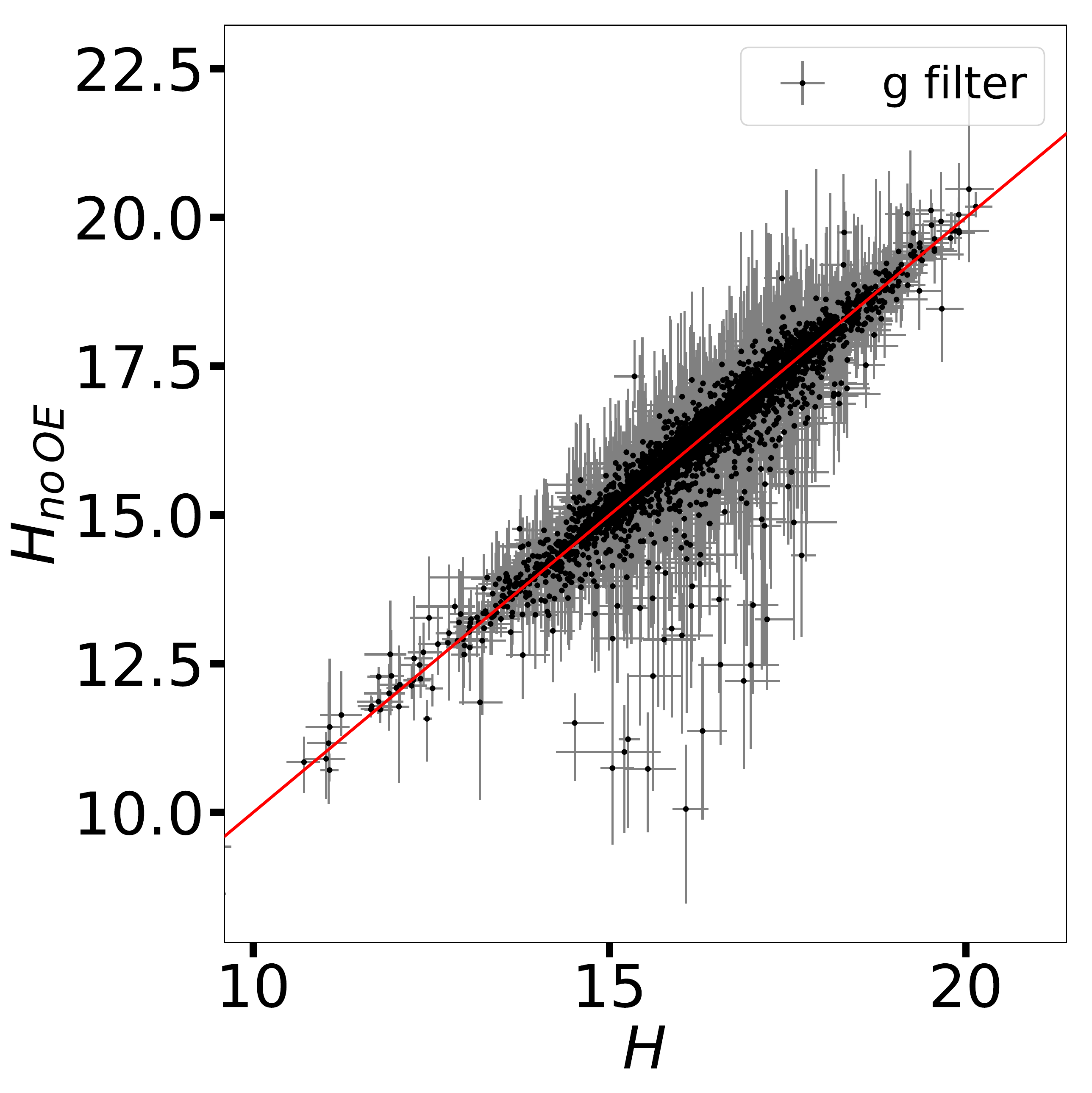}
 \includegraphics[width=6.5cm]{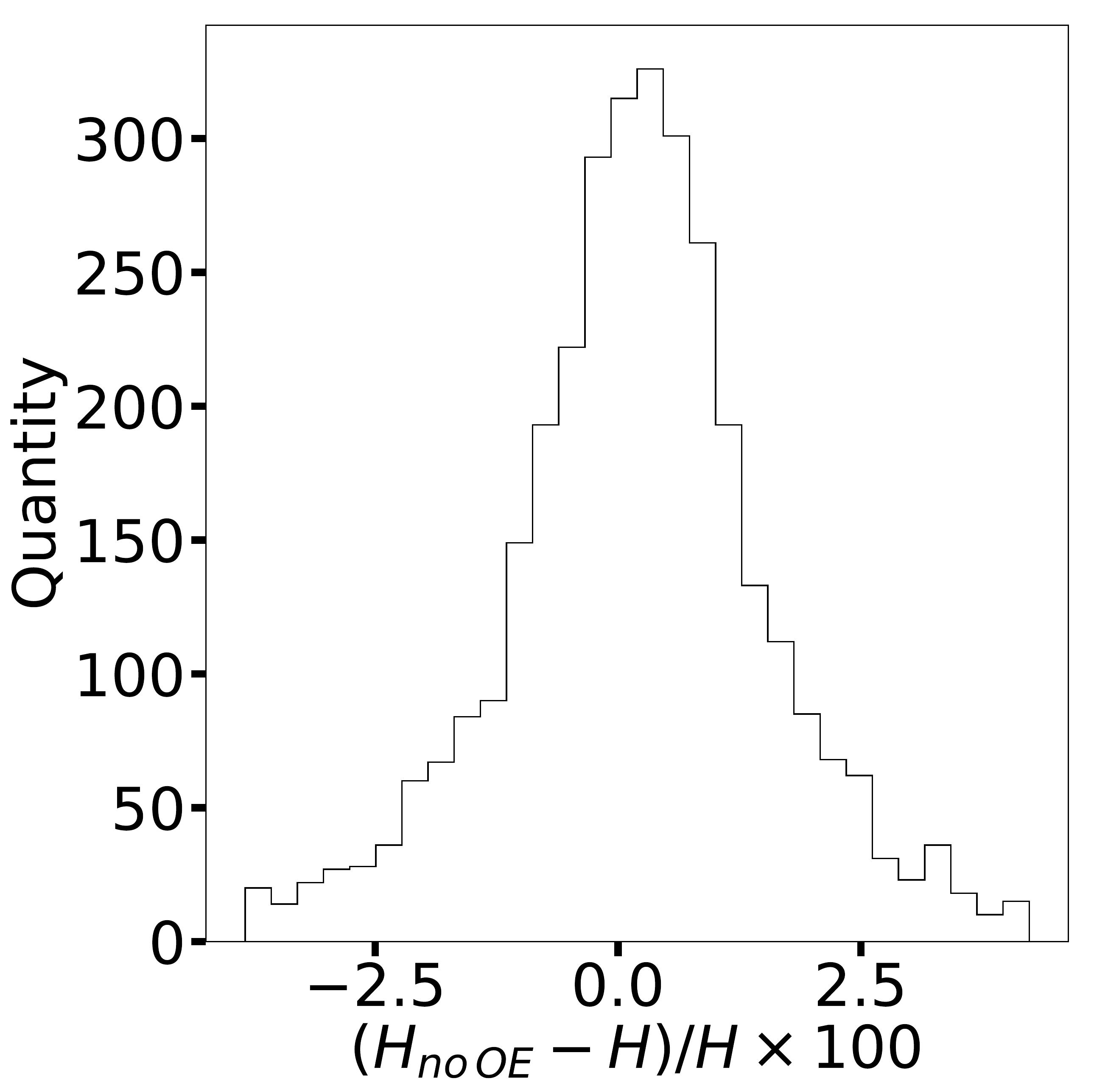}

\caption{Left panel: Comparison between $H_g$ obtained by us (labeled $H$) and the magnitude obtained eliminating all data with $\alpha<5$ degrees. The red line indicates the 1:1 relation. Right panel: Distribution of the percentual differences between the obtained $H$. The x-axis was cut for clarity.}\label{fig:noOE}%
\end{figure}

As seen in Fig. \ref{fig:fig3} (right panel), many of the objects analyzed in this work have $\alpha_{min}>5$ degrees; therefore, their data do not cover the highly non-linear region of the opposition effect (OE). That may account for systematic differences between $H$ obtained using PCs with data in the OE region with respect to data that do not. To test if this could affect our results, we run the code, using the HG$_{12}^*$ model but removing all data below 5 degrees. In this way, we can compare directly the same objects and check the impact of not covering the OE region. A comparison can be seen in Fig \ref{fig:noOE}. Numerical relevant values are shown in Table \ref{table:comparison}. Note that we restrict the comparison to the objects that had data removed; if initially, they did not have any data below 5 degrees, they were left unchanged.

\begin{figure}
\centering
\includegraphics[width=\hsize]{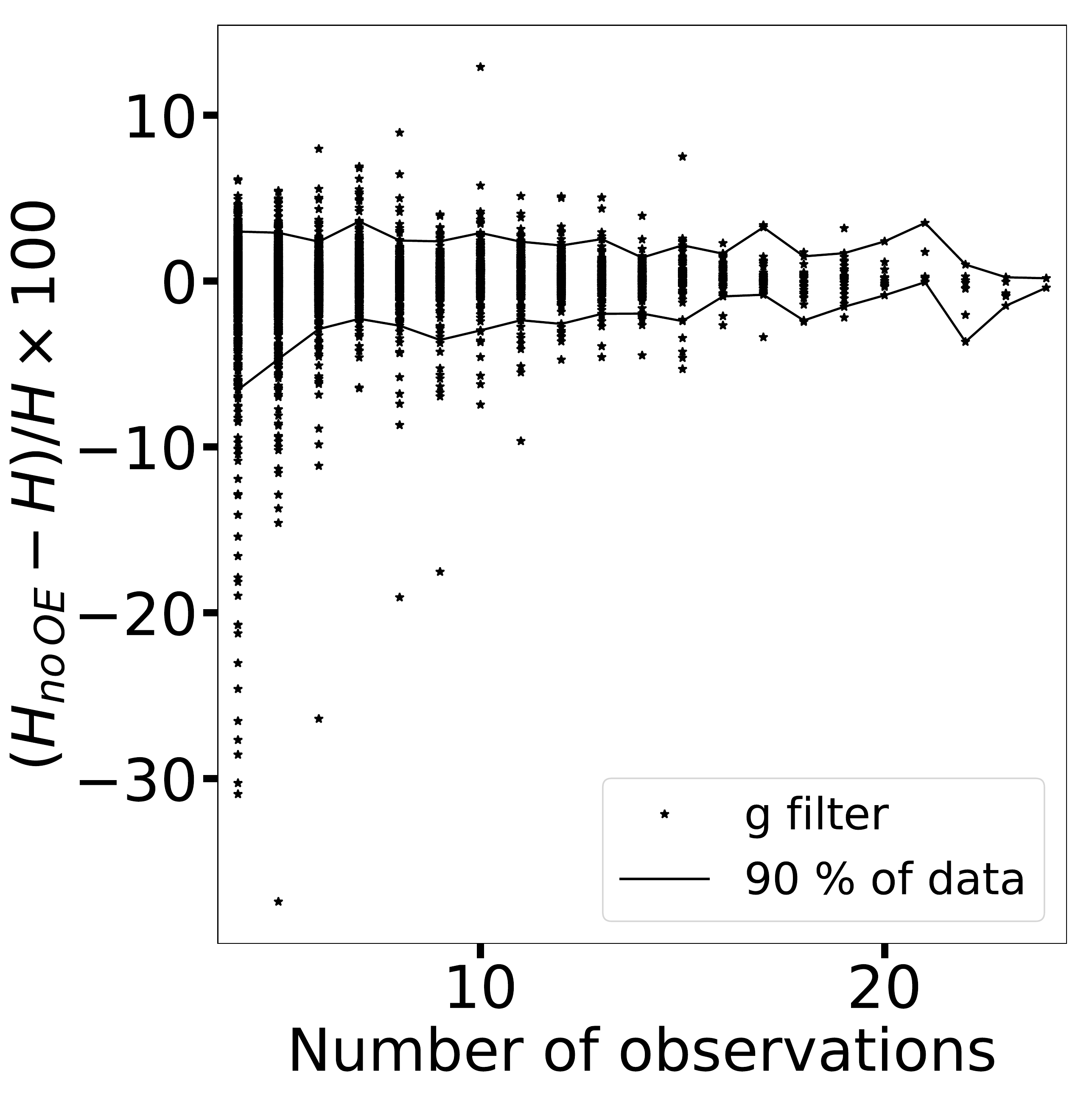}
\caption{Number of observations per percentile difference. The continuous lines enclose  the region where 90 \% of the solutions lie.}\label{fig:OEnumber}%
\end{figure}

The median values are close to zero. For instance, for an object with $H_g=16$, the median impact will account for a difference of 0.03 mag, well within our median uncertainties. Nevertheless, it is possible to see some substantial discrepancies, especially some objects that come out as brighter when the low-phase angle data is removed. These $H$s correspond to objects with a low number of observations, as shown in Fig. \ref{fig:OEnumber}. In the figure, the x-axis corresponds to the number of observations in the g filter before removal. 
The large differences seem to be related to significant uncertainties that arise when removing one or two points at low-phase angles from original data with four to ten input points. We do not detect any other strong systematic in our results. Therefore removing the OE region affects the uncertainty in the results when the original PC had few data points. However, we strongly remark that this does not happen for all objects. Therefore, we regard our results as trustworthy.

\subsection{Comparison with other works}
We compare our results with the four works mentioned above: O11, V15, C21, and M21. The former work presented over 400\,000 PCs of objects using data compiled from the Minor Planets Center. Initially very heterogeneous in observatories, instruments, and filters, these data were carefully calibrated to a V filter magnitude to be analyzed. In V15 were presented almost a quarter of a million $H_V$ from the Pan-STARRS data release 1. The magnitudes were computed using Monte Carlo techniques to include effects of rotational variations and taxonomy heterogeneity. The data were magnitudes obtained in SLOAN-like filters that were transformed to $V$. With this comparison, we aim to check the quality of our results with a well-tested sample. In the case of C21, they used data from the Gaia DR2. In particular, we will compare with the sample of objects observed {\it only} with the Gaia G filter ($\gtrsim14\,000$ objects), without transforming it to $V$. Due to the scanning law used by Gaia, these data were taken with $\alpha>10$ degrees. This comparison aims to see how our data compare with one of the largest databases of small bodies photometric measurements. Finally, M21 presented 127\,012 PCs for 94\,777 asteroids in a two-filter system: cyan and orange. The absolute magnitudes and phase coefficients were obtained using Bayesian inference and a Monte Carlo approach and reported as the mean values of the posterior distributions and the 95 \% high-density interval, i.e., the interval where lie 95 \% of the solutions.

The first three works used the HG$_{12}$ system as presented in \cite{muinonen2010HG1G2} and not the HG$_{12}^*$ version presented in \cite{penti2016HG} (O11 and V15 came out before it was published), while M21 did use the HG$_{12}^*$. Keep in mind that $G_{12}$ and $G_{12}^*$ are not the same quantities. Due to this heterogeneity, we focus our comparison on the absolute magnitudes and not the phase coefficients, although, as shown below, we can draw useful information from the distribution of uncertainties of the different catalogs.

\begin{figure}
\centering
 \includegraphics[width=6.5cm]{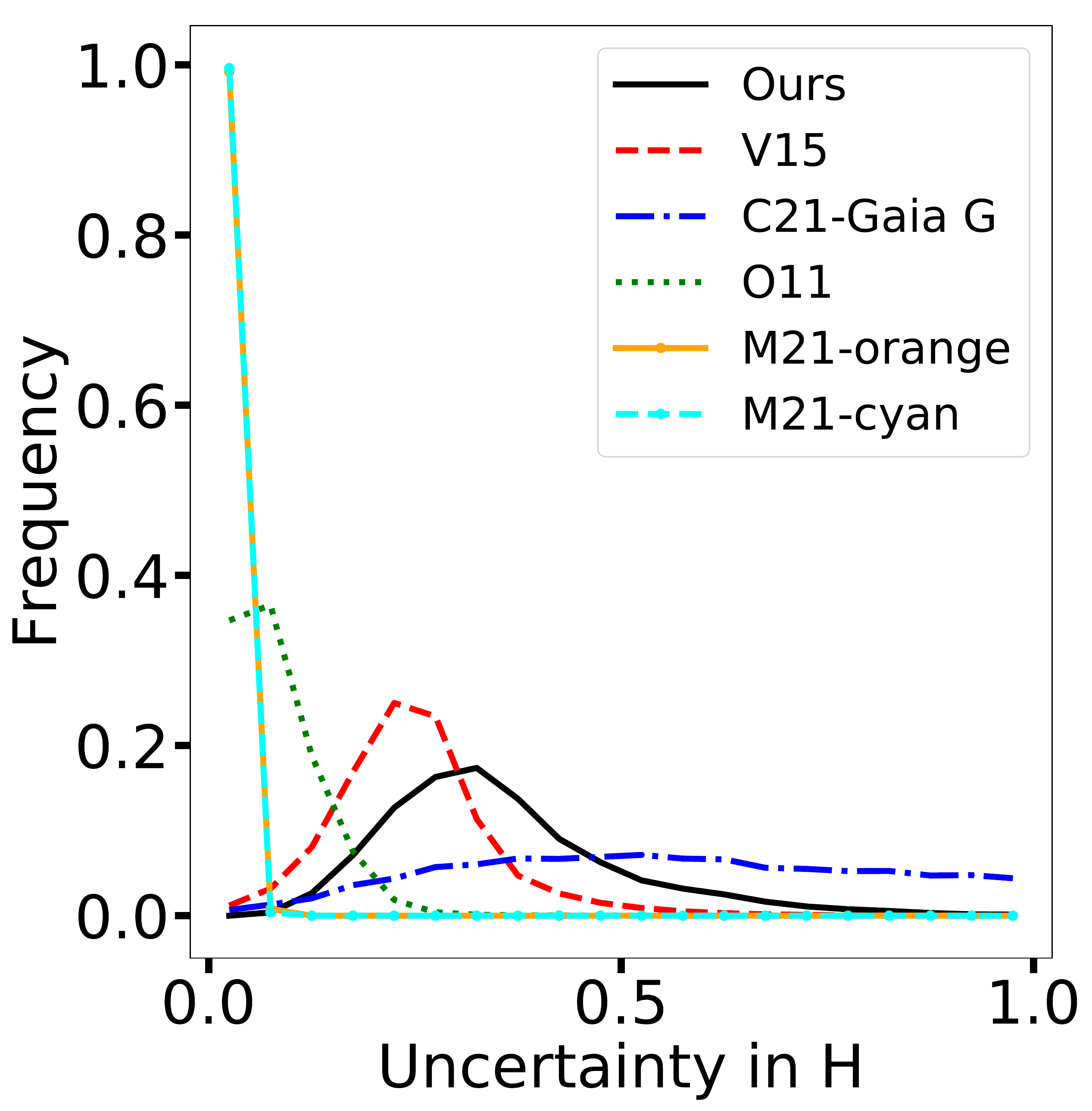}
 \includegraphics[width=6.5cm]{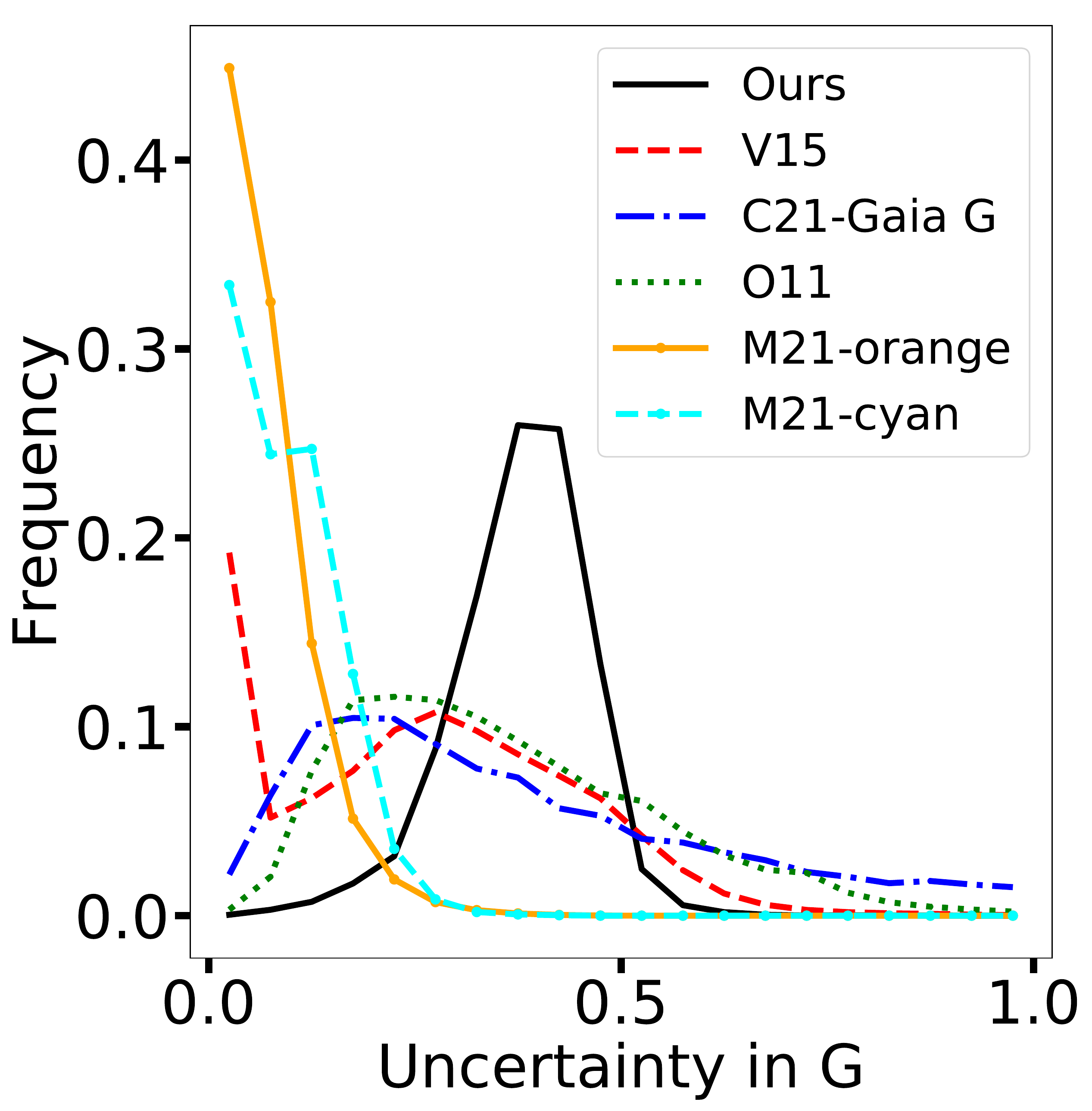}

\caption{Comparison of the frequency distributions of the uncertainties in $H$ (left panel) and $G$ (right panel) between this work (black continuous line), O11 (green dotted line), V15 (red dashed line), $H_G$ of C21 (blue dot-dashed), $H_{cyan}$ (cyan dashed curve with big dots), and $H_{orange}$ (orange continuous line with big dots), the last two from M21. The x-axes were arbitrarily cut at 1 to exclude large values. }\label{fig:fig5}%
\end{figure}

\begin{figure}
\centering
 \includegraphics[width=6.5cm]{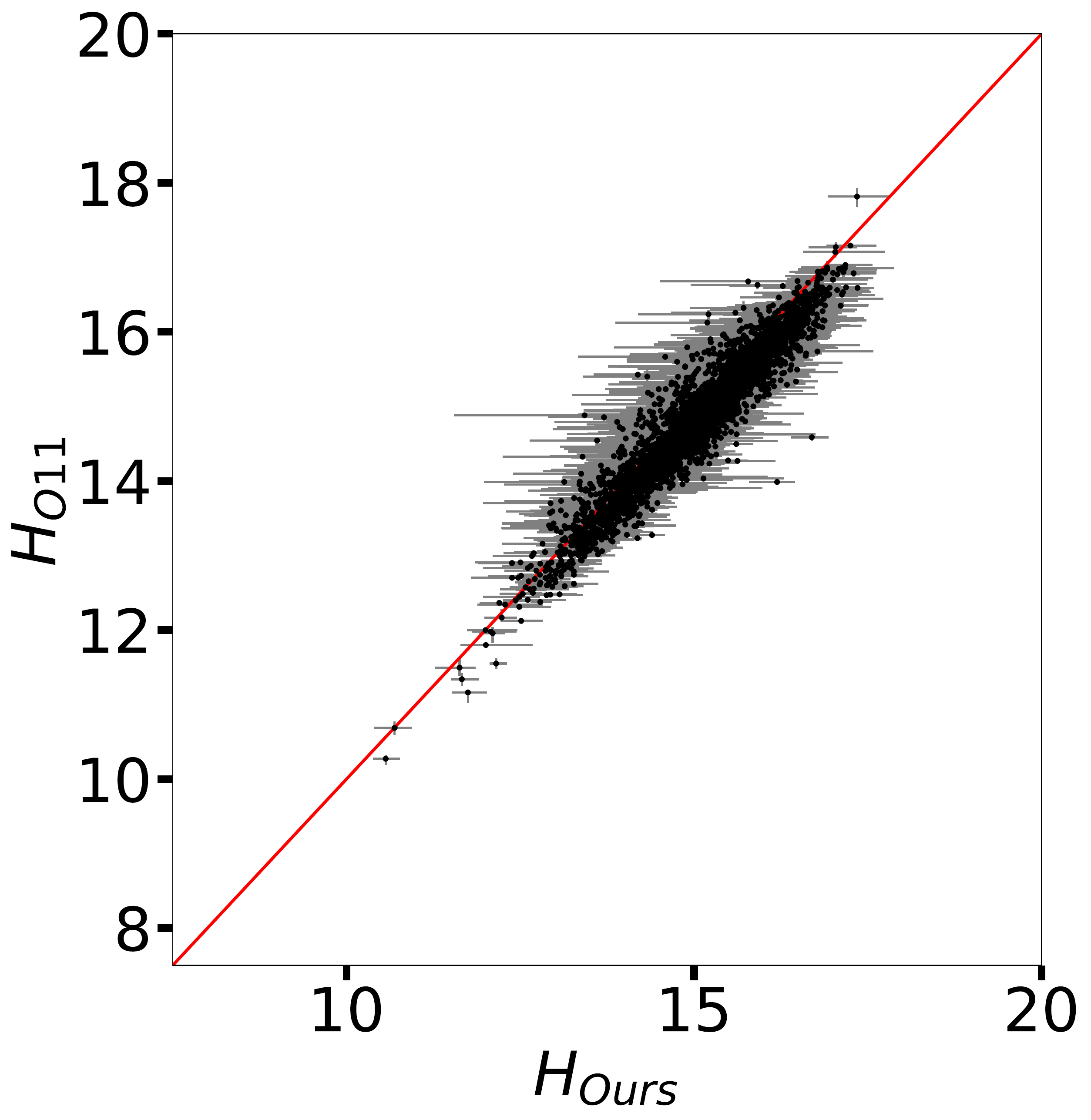}
 \includegraphics[width=6.5cm]{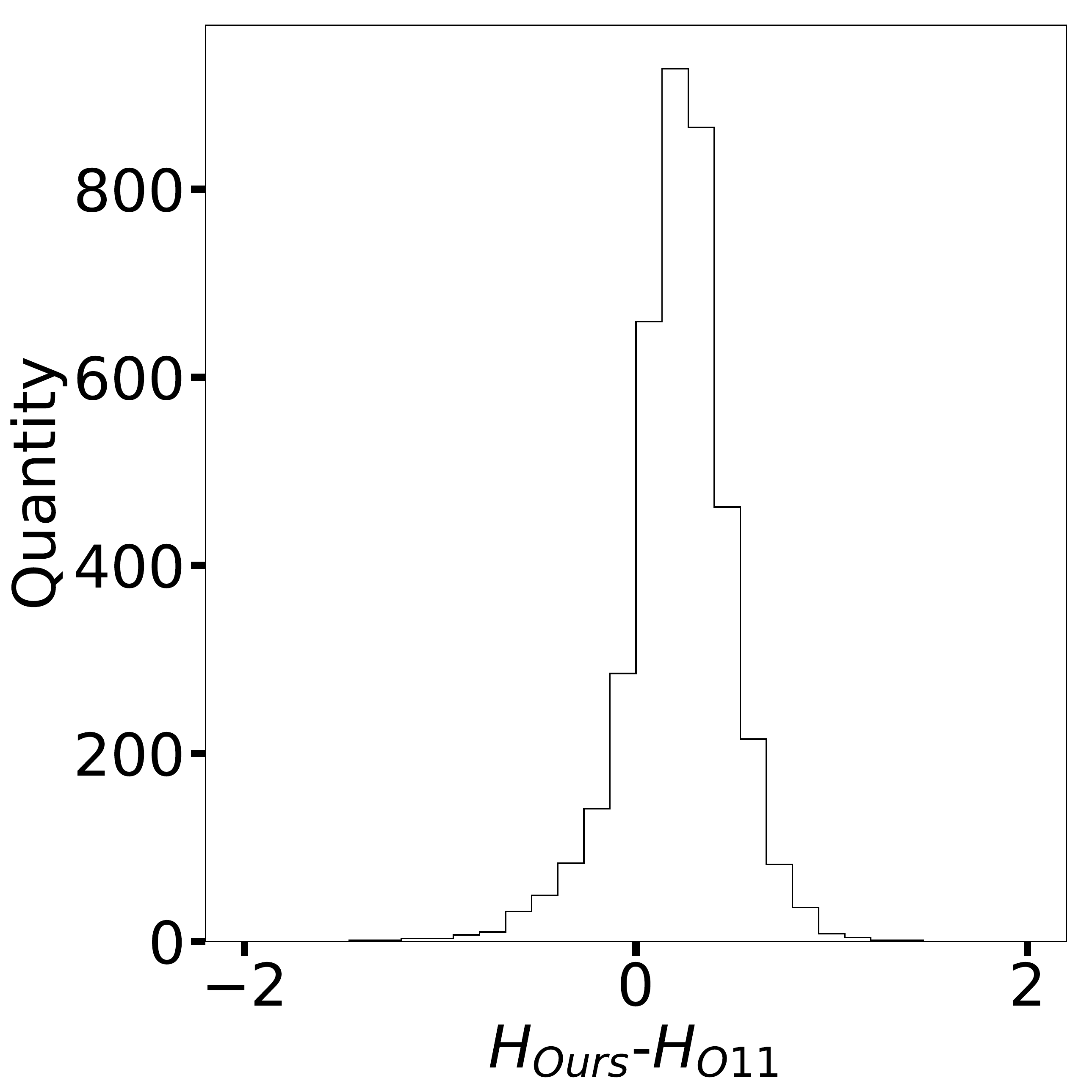}
\caption{Left panel: Comparison between $H_V$ obtained by us and in O11. The red line indicates the 1:1 relation. Right panel: Distribution of the differences between the obtained $H$.}\label{fig:fig12}%
\end{figure}

\begin{figure}
\centering
 \includegraphics[width=6.5cm]{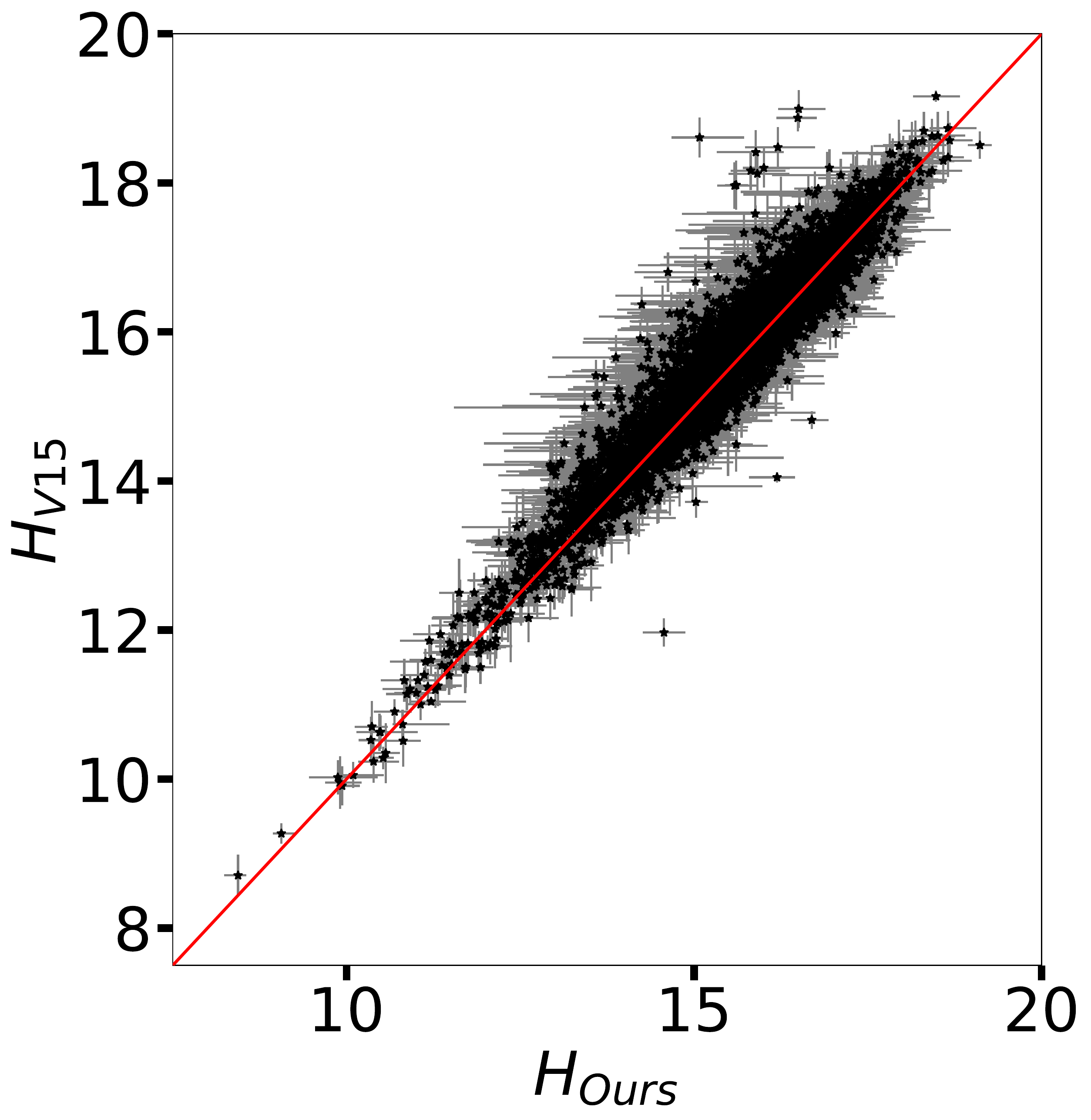}
 \includegraphics[width=6.5cm]{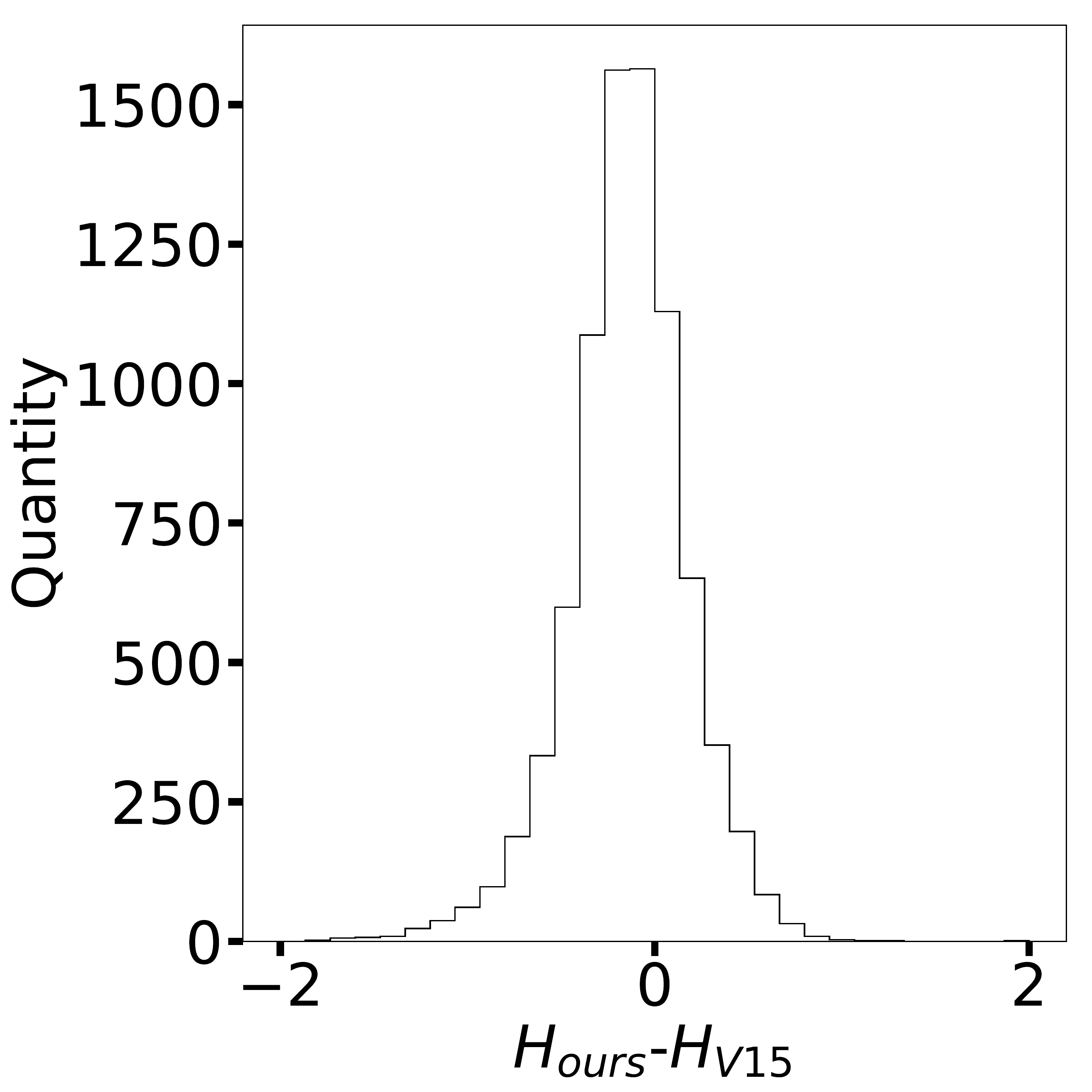}

\caption{Left panel: Comparison between $H_V$ obtained by us and in V15. The red line indicates the 1:1 relation. Right panel: Distribution of the differences between the obtained $H$.}\label{fig:fig6}%
\end{figure}

\begin{figure}
\centering
 \includegraphics[width=6.5cm]{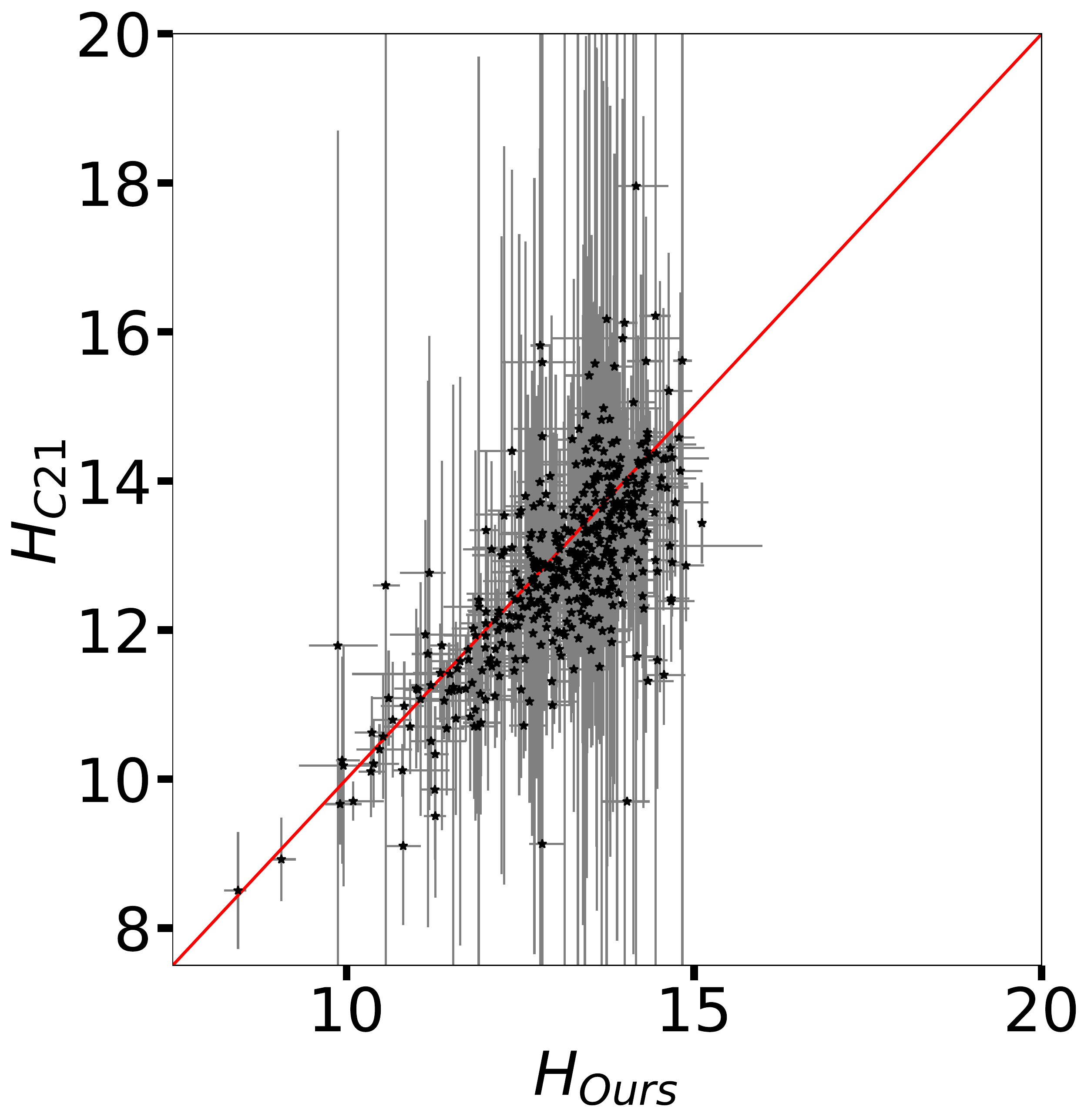}
 \includegraphics[width=6.5cm]{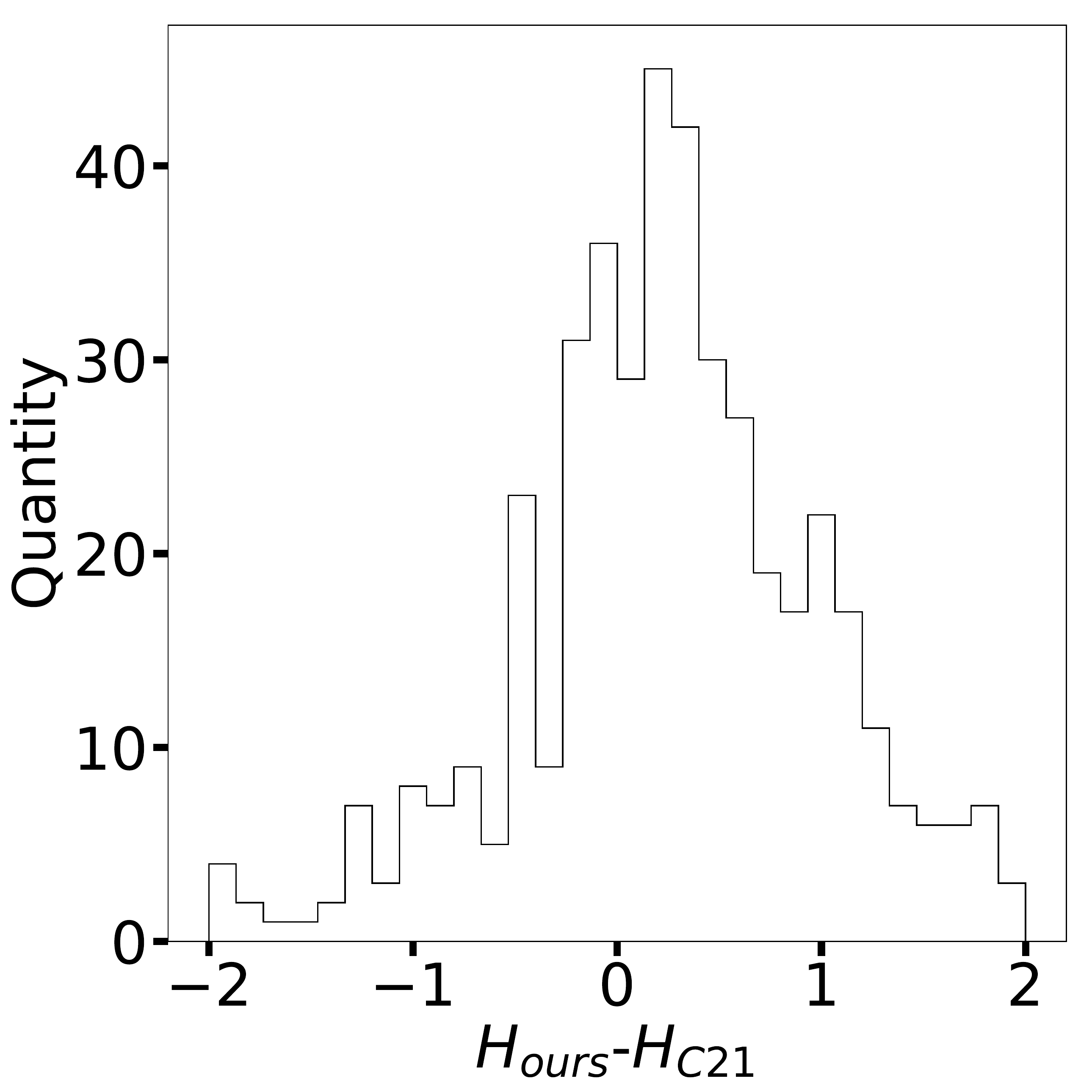}
\caption{Left panel: Comparison between $H_V$ obtained by us and $H_G$ obtained by C21. The red line indicates the 1:1 relation. Right panel: Distribution of the differences between the obtained $H$.}\label{fig:fig7}%
\end{figure}

\begin{figure}
\centering
 \includegraphics[width=6.5cm]{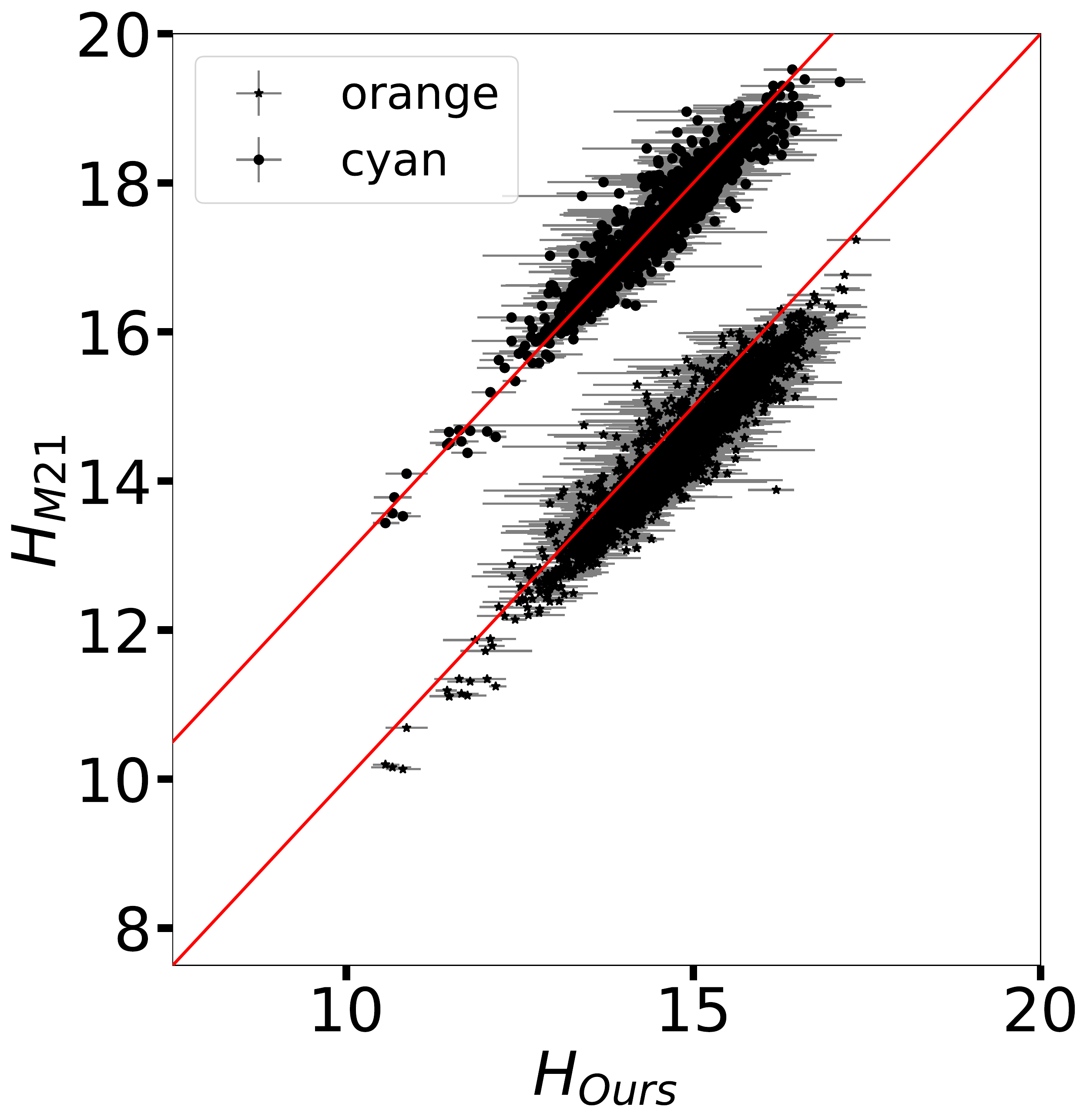}
 \includegraphics[width=6.5cm]{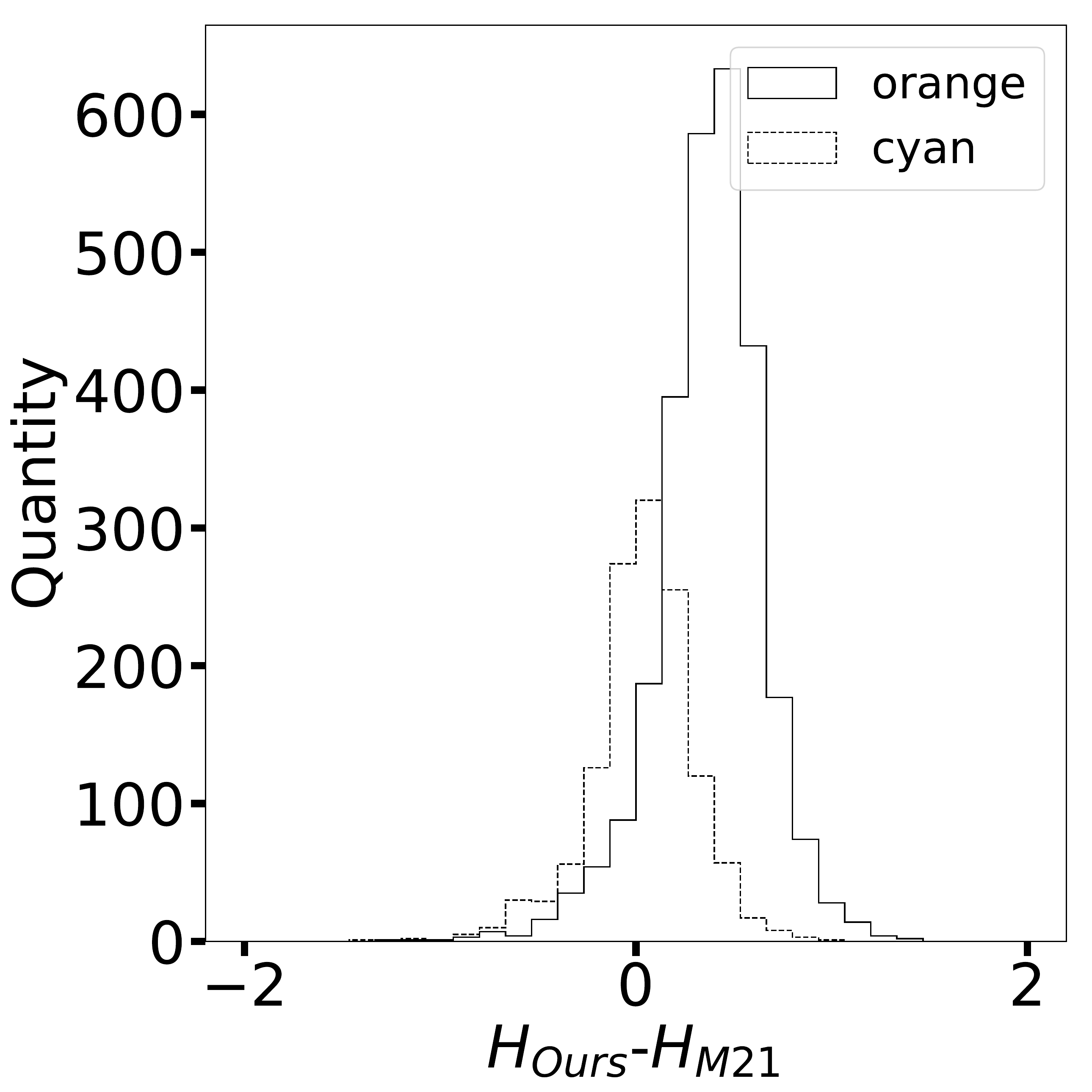}
\caption{Left panel: Comparison between $H_V$ obtained by us and $Hs$ obtained by M21. The red line indicates the 1:1 relation. The comparison with the cyan magnitudes were shifted by 3 in the y-axis for clarity. Right panel: Distribution of the differences between the obtained $H$.}\label{fig:Ours_M21}%
\end{figure}

Figures \ref{fig:fig5} to \ref{fig:Ours_M21} show the comparison of the $H$ values from all researches ($H_V$ from this work, O11, and V15, $H_G$ in C21, and $H_{cyan/orange}$ in M21). C21 reached a brighter population than the other four works (see the numerical comparison in Table \ref{table:3}). The difference cannot be explained only by the difference in the filters used: For an object with the color of the Sun $(V-R)=0.352$ \citep[from][]{ramirez2012} and using the transformations from Gaia system to Johnson-Cousins\footnote{https://gea.esac.esa.int/archive/documentation/GDR2/Data\_processing/chap\_cu5pho/sec\_cu5pho\_calibr/ssec\_cu5pho\_PhotTransf.html} we obtain $(G-V)=-0.13$. Interestingly, this result explains half the difference between our $H$ and C21's (see Table \ref{table:3}). M21's magnitudes are also brighter than our $H_V$, and these of O11 and V15, although not as much as C21.  On the other hand, our distribution of $H_V$ is very similar, though slightly brighter but still within one standard deviation of the distributions ($\gtrsim1$ mag), to O11's and V15's. 
\begin{table*}
\caption{Comparison between this work, O11, V15, C21, and M21.}
\label{table:3}
\centering
\begin{tabular}{c c | c c}
\hline\hline
             & Median\tablefootmark{a} &  & Median (standard deviation)\tablefootmark{b}\\
\hline
$H_V$ (Ours)        & 16.0              & $H_V$ (Ours) - $H_V$ (O11) &  0.23 (0.27) \\
$H_V$ (O11)         & 16.0              & $H_V$ (Ours) - $H_V$ (V15) & -0.13 (0.33) \\
$H_V$ (V15)         & 16.1              & $H_V$ (Ours) - $H_G$ (C21) &  0.26 (0.89) \\
$H_G$ (C21)         & 12.9              & $H_V$ (Ours) - $H_{orange}$ (C21) & 0.40 (0.28) \\
$H_{orange}$ (M21)  & 14.6              & $H_V$ (Ours) - $H_{cyan}$ (C21)   & 0.06 (0.27) \\
$H_{orange}$ (M21)  & 14.4                      &                            &            \\
\hline\hline
             & Median Uncertainty\tablefootmark{a} &  & Median Uncertainty\tablefootmark{a}\\
\hline
$H_V$ (Ours)          & 0.33& $G_V$          (Ours)  & 0.39 \\
$H_V$ (O11)           & 0.07& $G_{12V}$      (O11)   & 0.33 \\
$H_V$ (V15)           & 0.24& $G_{12V}$      (V15)   & 0.26 \\
$H_G$ (C21)           & 0.98& $G_{12G}$ (C21)        & 0.55 \\
$H_{orange}$ (M21)    & 0.01& $G^*_{12orange}$ (M21) & 0.06 \\
$H_{cyan}$ (M21)      & 0.02& $G^*_{12cyan}$ (M21)   & 0.08 \\

\hline
\end{tabular}
\tablefoot{
{\tablefoottext{a}{Whole valid sample.}}
{\tablefoottext{b}{Common sample.}}}
\end{table*}

We also compared the uncertainties in the five works. Note that we defined the uncertainties as to the interval edges where 68\% of the solutions lie. This gives asymmetric quantities when referred to the median of the distribution. In order to compare our uncertainties, we chose as the metric the uncertainty defined as $\{[(\sigma_{H_V}^+)^2+(\sigma_{H_V}^-)^2]\times0.5\}^{0.5}$. The uncertainties in O11 were treated similarly. In M21 are not reported uncertainties, but instead, magnitudes corresponding to the limits of the 95 \% high-density interval. We will consider this interval as the formal uncertainty. It is clear from Fig. \ref{fig:fig5} that M21 and O11 have very precise values of $H_V$, with uncertainties lower than the other three\footnote{Keep in mind that precision has nothing to do with how close the result is to the real value (accuracy), but with the quality of the measurement.}. Most of M21's uncertainties, in both filters, lie within the first bin in the figure that is the reason for the abrupt behavior of the curve. Meanwhile, C21 has a long tail of high-uncertainty $H$ and $G$. Our data seem to have larger uncertainties than O11, V15, and M21. However, they do not have the considerable uncertainties that appear for C21, especially in $G$. The long tail of highly uncertain values seen in C21 is possibly related to the fact that their data do not reach $\alpha<10$ degrees (see also Table \ref{table:3}). The authors complemented their PCs with ground-based $V$ magnitudes (previously transforming Gaia $G$ magnitude to $V$) to cover $\alpha<10$ degrees. Unfortunately, there are only six objects in common with our results, therefore precluding any reliable comparison. 

So far, we have compared the whole valid datasets (note that, for example, in V15's online catalog, there are lines with negative uncertainties, which we do not consider valid). Now we compare the objects that are in common. First, we compare our results with O11, then with V15 and C21, and finally with M21 (significant numbers are already displayed in Table \ref{table:3}). As mentioned above, it is not possible to compare the values of $G$ because they were obtained using different definitions. Nevertheless, we can compare the absolute magnitudes obtained.

Our first comparison is with O11's $H_V$ (Fig. \ref{fig:fig12}). The red line over-plotted corresponds to a 1:1 correlation. The comparison is made over 3\,879 objects in common $H_V\in(10,18)$. The agreement is good between both estimates of the absolute magnitudes, being ours fainter than O11's with a median difference of 0.23 mag, while most objects have a difference in $H_V$ between 0 and 0.5 mag. We consider the agreement to be quite good.
Figure \ref{fig:fig6} (left panel) shows the comparison between $H_V$ from this work and those published in V15.  The figure shows that there is a very good agreement between both sets (8\,050 objects in common), being our magnitudes slightly brighter than V15's (median difference -0.13 mag), while most of the objects lie between a range of 1 mag centered in 0 (right panel of Fig. \ref{fig:fig6}).
Next, we compare with a less dense database than the others, but with data that will probably become a benchmark of PCs studies. We compare with Gaia DR2 data as presented by \cite{colazo2021} (Fig. \ref{fig:fig7}). There are only 521 objects in common (some of them with non-valid results). Let us stress that their results were obtained using the G filter of Gaia, which is very broad when compared with the V filter. Overall, the data seems to follow the same trend, being C21's magnitudes brighter than ours (median difference 0.26 mag), which makes sense considering the broader nature of the Gaia G filter. In this case, it is possible to see more spread in the difference of magnitudes, while most of the objects lie $H_{C21}-H_{ours}\in(-0.5,1.0)$.
Finally, we show the comparison with the ATLAS database \citep{mahlke2021}. These data were obtained in two filters: cyan and orange, which are broader than the V filter. We have 1\,316 objects in common with the cyan magnitudes and 2\,743 with the orange magnitudes (Fig. \ref{fig:Ours_M21}). The comparison is quite good in both cases, especially with the cyan magnitudes (median difference 0.06), while our $H_V$ tends to be fainter than M21's (median difference about 0.40). This may be because the cyan filter covers most of the V filter, while the orange filter only covers the reddest part of the V filter transmission curve.

\subsection{Zero phase angle colors}\label{colors}
\begin{figure}
\centering
 \includegraphics[width=6.5cm]{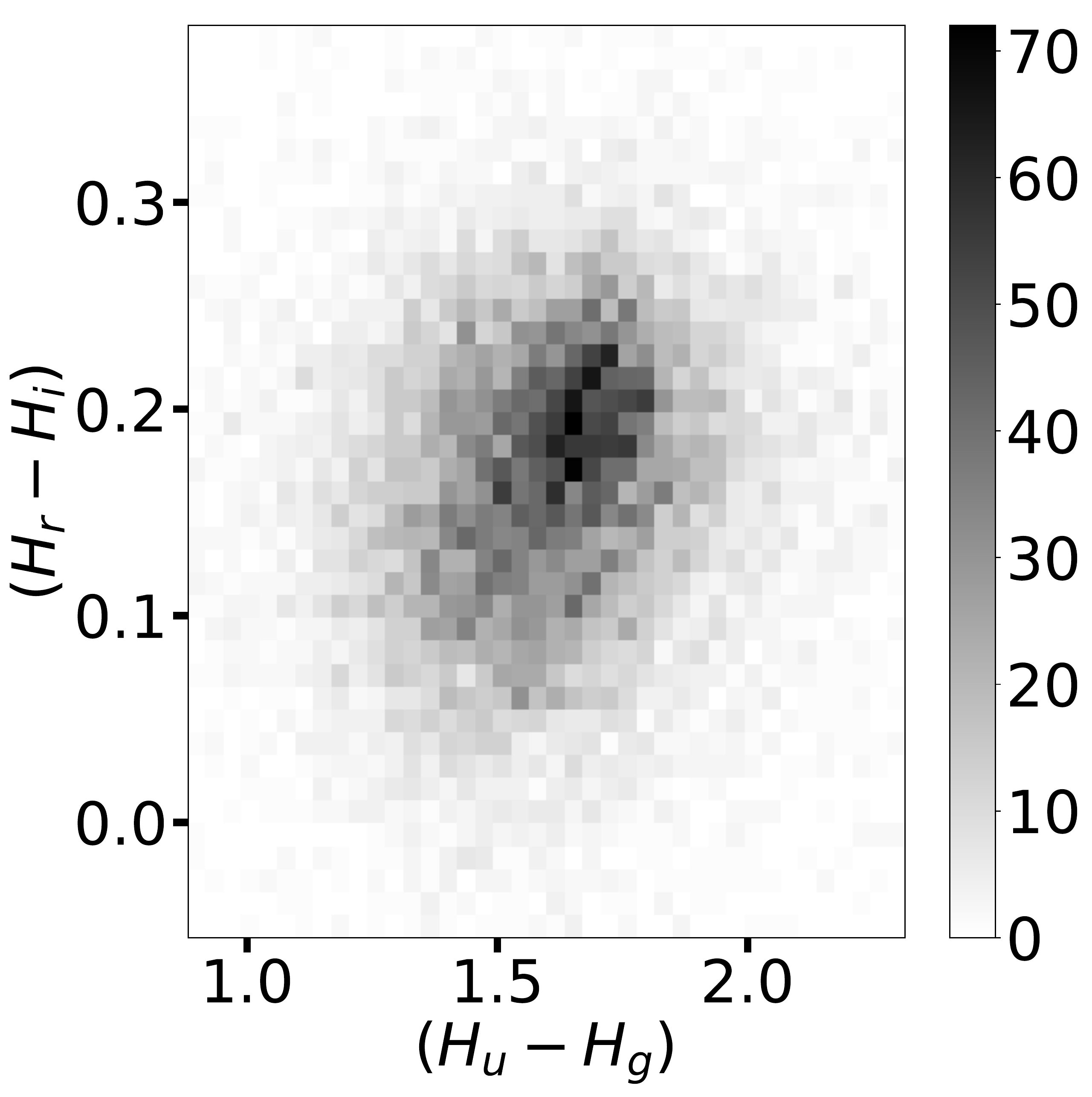}
 \includegraphics[width=6.5cm]{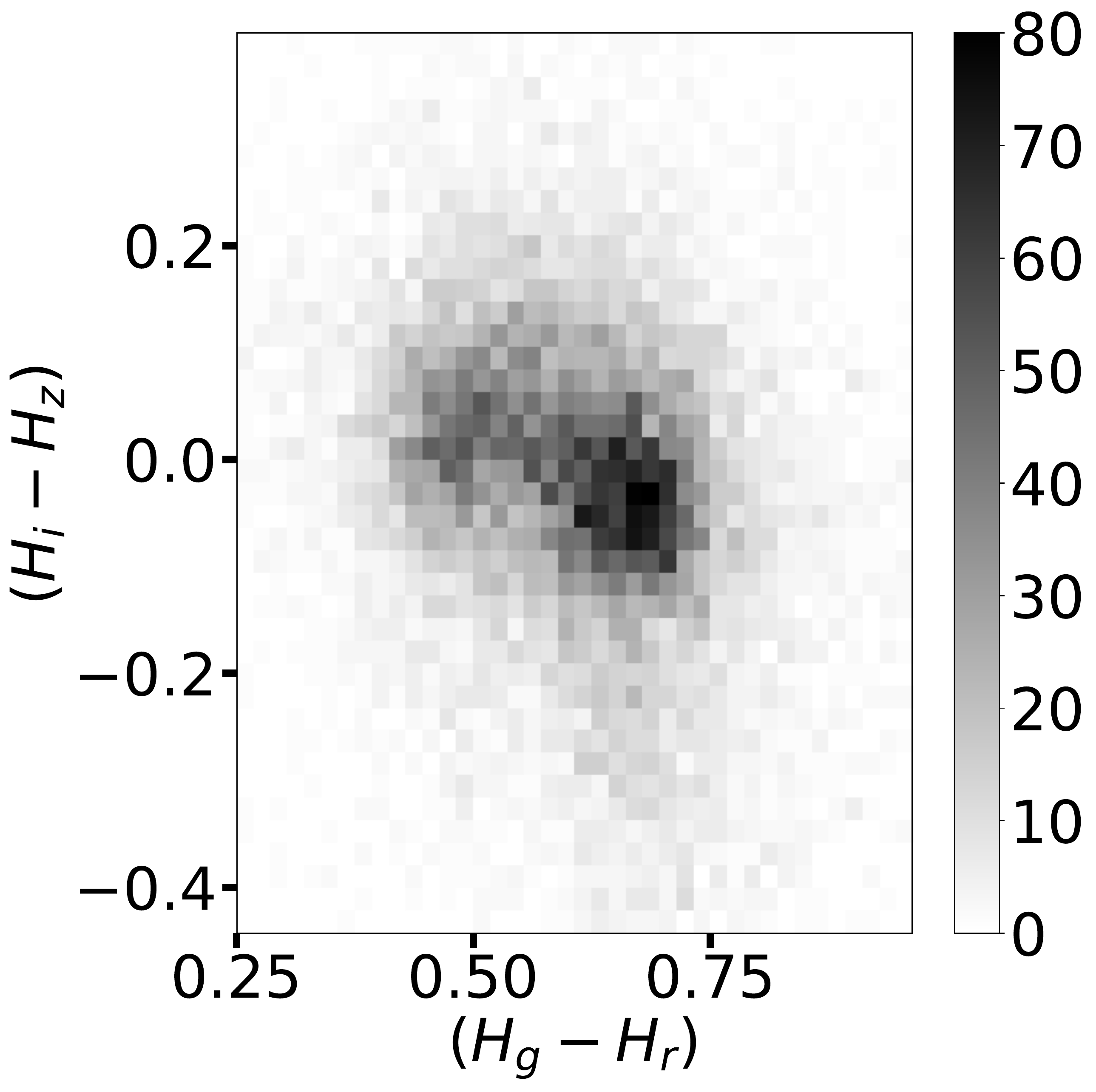}
\caption{Absolute color - color diagrams. The axes were cut to exclude a few remaining outliers.}\label{fig:fig11}%
\end{figure}

Following \cite{ayala2018} we define the absolute colors as the difference in the absolute magnitudes in different filters. These colors {\it are not} affected by phase coloring and therefore represent a template to study these effects. The colors are shown as heat diagrams in Fig. \ref{fig:fig11}. A quick comparison with Fig. 8 in \cite{ivezic2001AJ} shows that the space covered is similar (but logically lower in numbers).

In this work, we do not attempt to study the possible taxonomical distribution of our objects (as done in \citealt{Oszki2012Icar}), nor its distributions. These issues will be addressed in a future work. Nevertheless, it is clear, especially looking at the right panel of Fig. \ref{fig:fig11}, that we do recover the large bimodal distribution of colors, historically representing the C and S type asteroids (see, for example, \citealt{hasselmann2015}) while also seeing a tail of object going down to $(H_g-H_r)\approx0.7$ and $(H_i-H_z)<0$ associated to V-type asteroids.

\subsection{Absolute colors vs. average colors}\label{absvsave}
\begin{figure}
\centering
 \includegraphics[width=6.5cm]{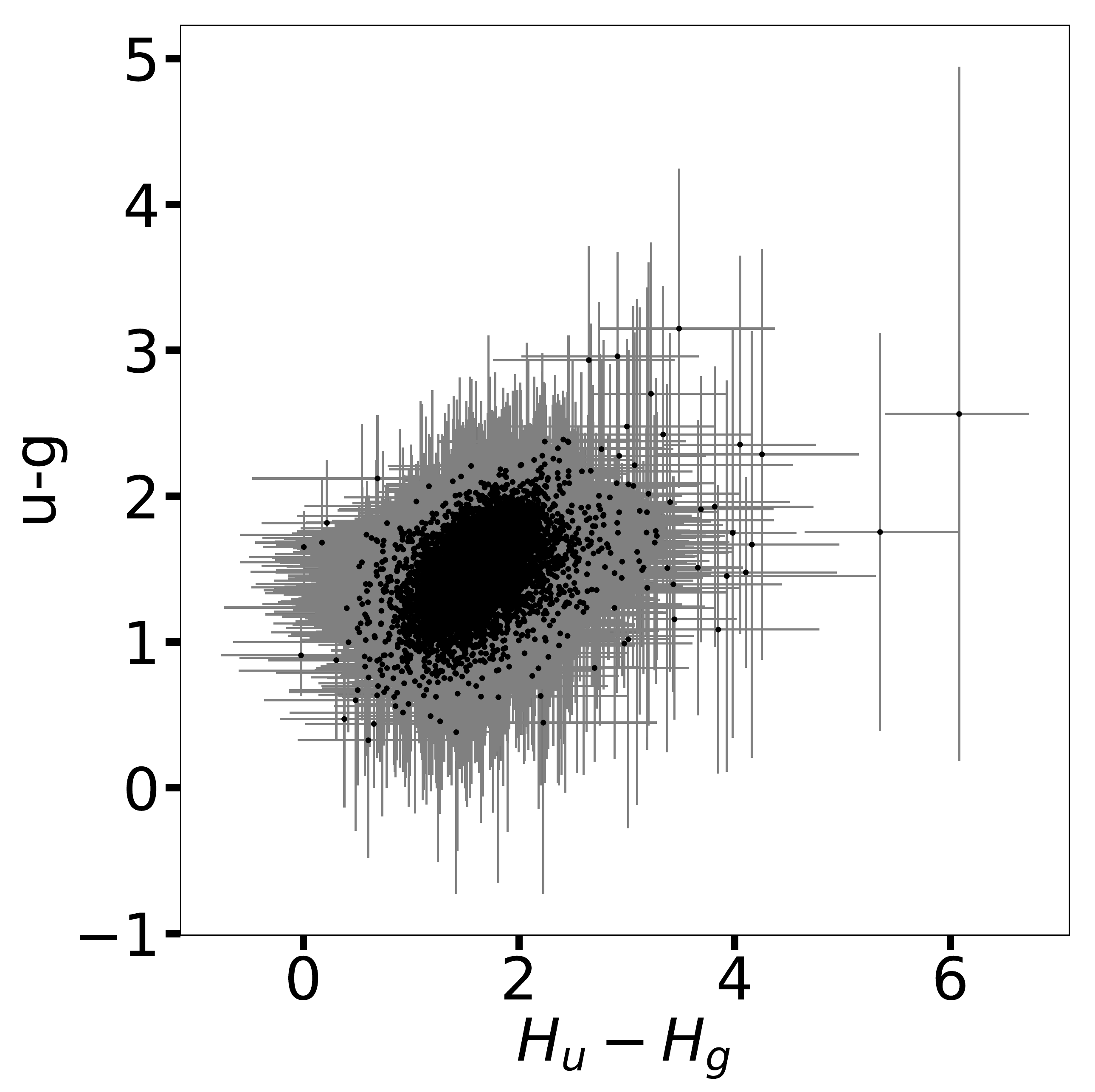}
 \includegraphics[width=6.5cm]{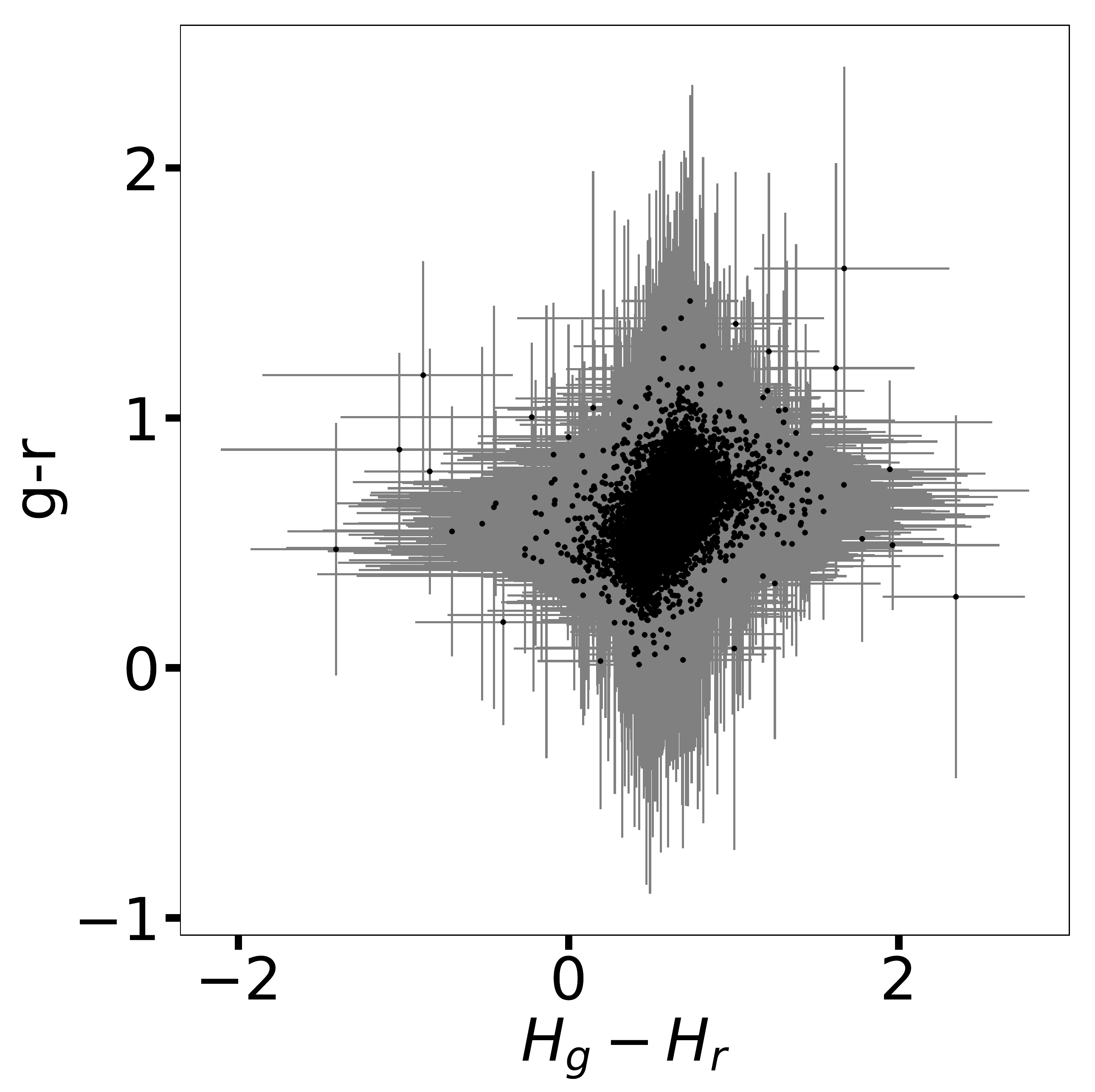}
 \includegraphics[width=6.5cm]{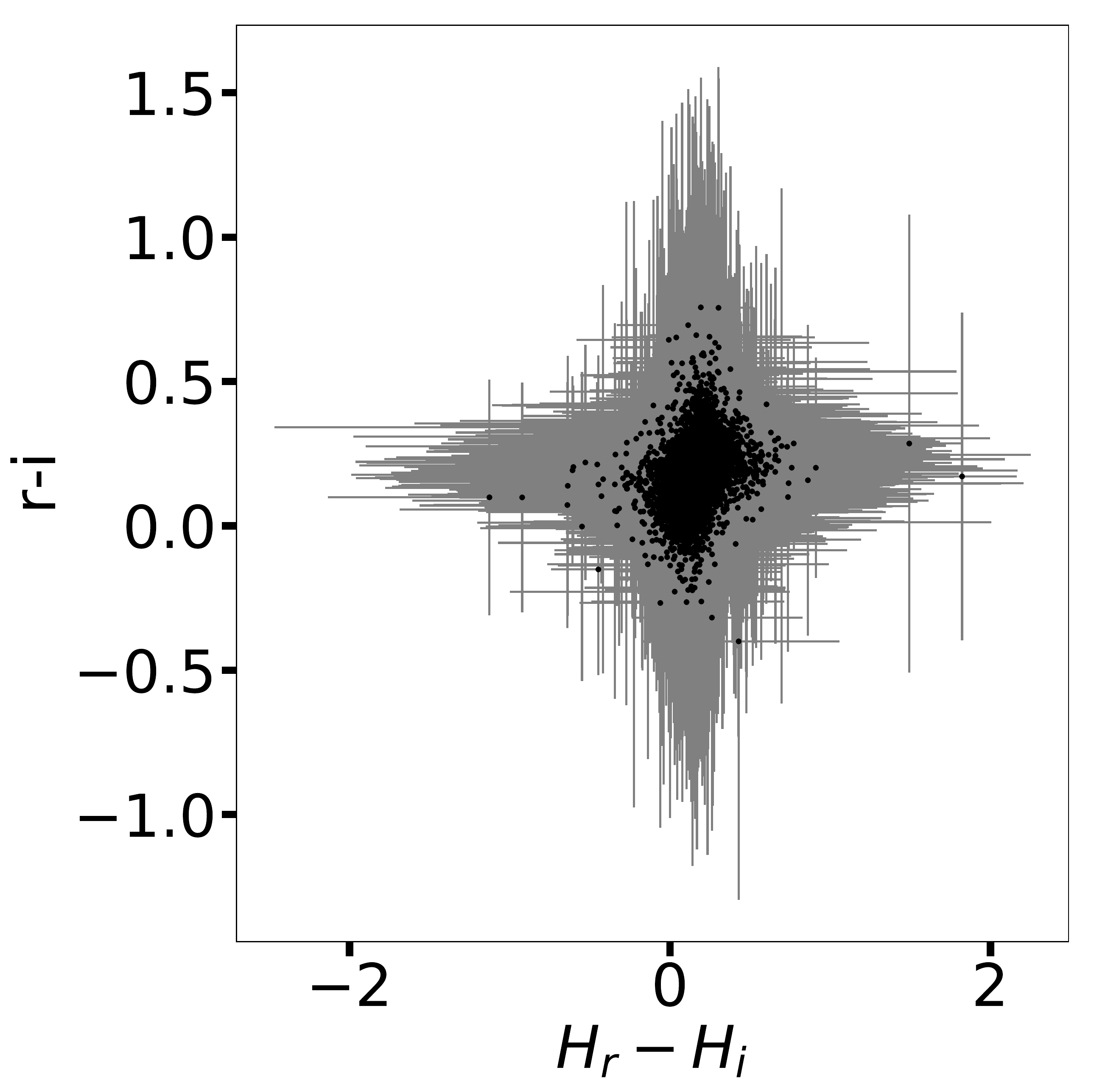}
 \includegraphics[width=6.5cm]{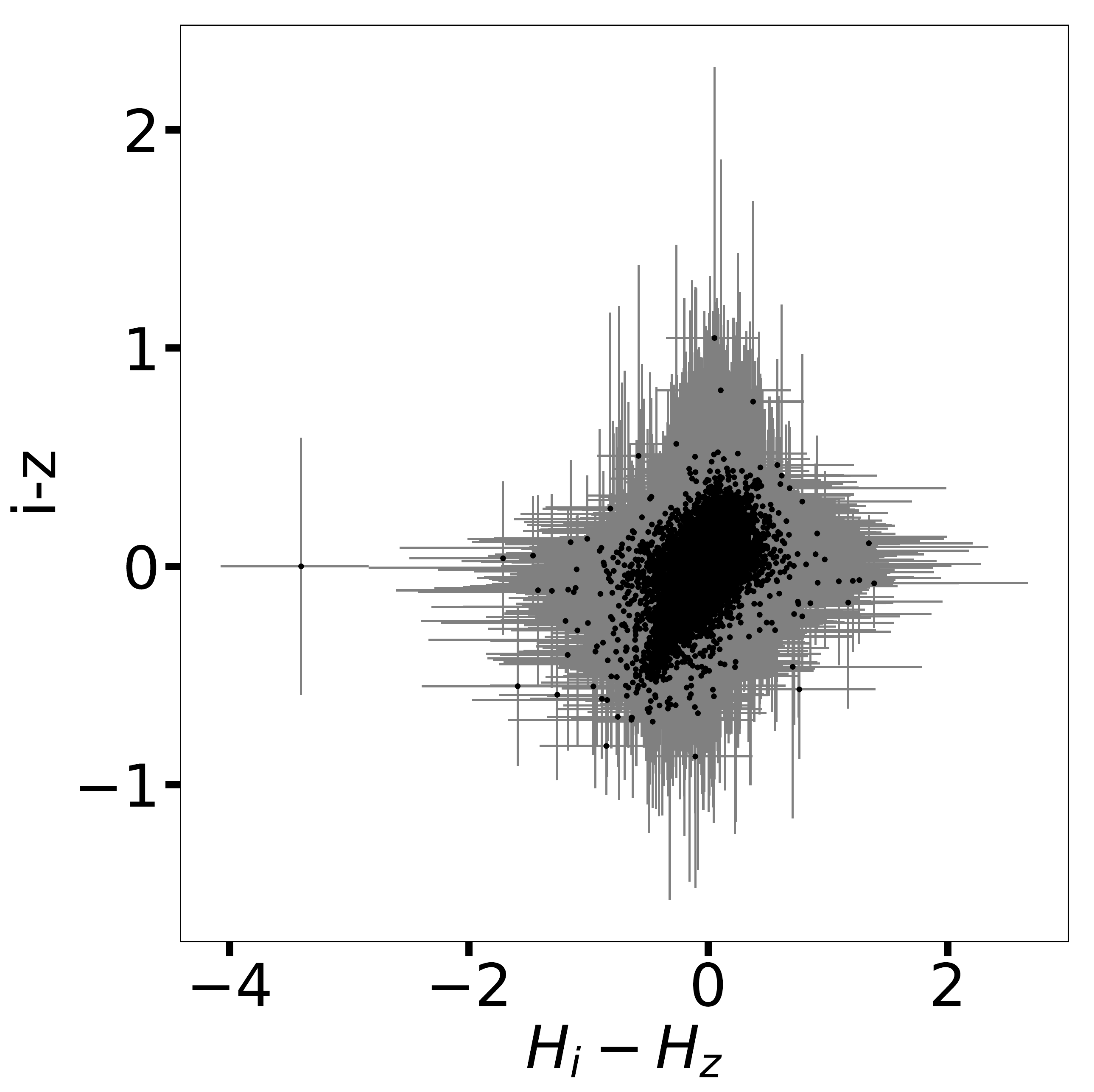}

\caption{Absolute colors vs. average colors: Top left, $u-g$; top right, $g-r$; bottom left, $r-i$; bottom right, $i-z$. The typical (median) error bars are shown separately for clarity.}\label{fig:Absvsave}%
\end{figure}

One of our motivations is to produce a series of data that are not affected by phase coloring and can, therefore, be used with confidence to study, for example, the taxonomical distribution of objects in the main belt of asteroids. In Fig. \ref{fig:Absvsave} we show the weighted average of the colors per object compared with their respective absolute colors. These weighted colors were computed as the weighted average of the same data used to create the PCs. The weights are $1/\sigma_m^2$. The weighted standard deviation gives the error in each average color. Both sets of colors correlate (in all cases, the Spearman test produced $r_s \gtrsim 0.5$ and $P_{r_s}\approx0$), but it is clear that there is not a 1 to 1 relation, reinforcing the idea that the average colors, although useful, do not necessarily represent the actual colors of the object and should be interpreted with great care.

\subsection{Absolute colors and $\Delta G$}
\begin{figure}
\centering
 \includegraphics[width=6.5cm]{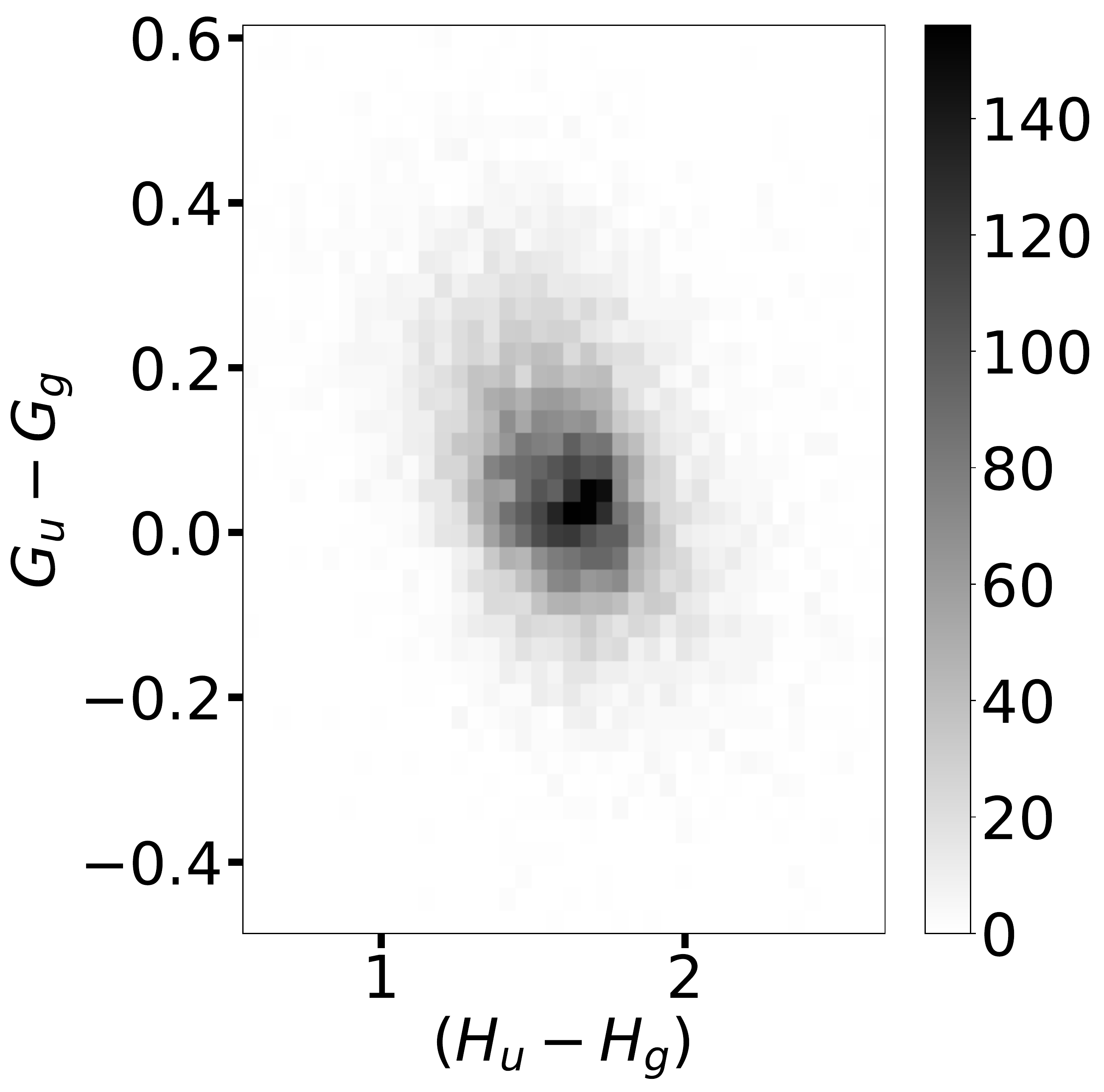}
 \includegraphics[width=6.5cm]{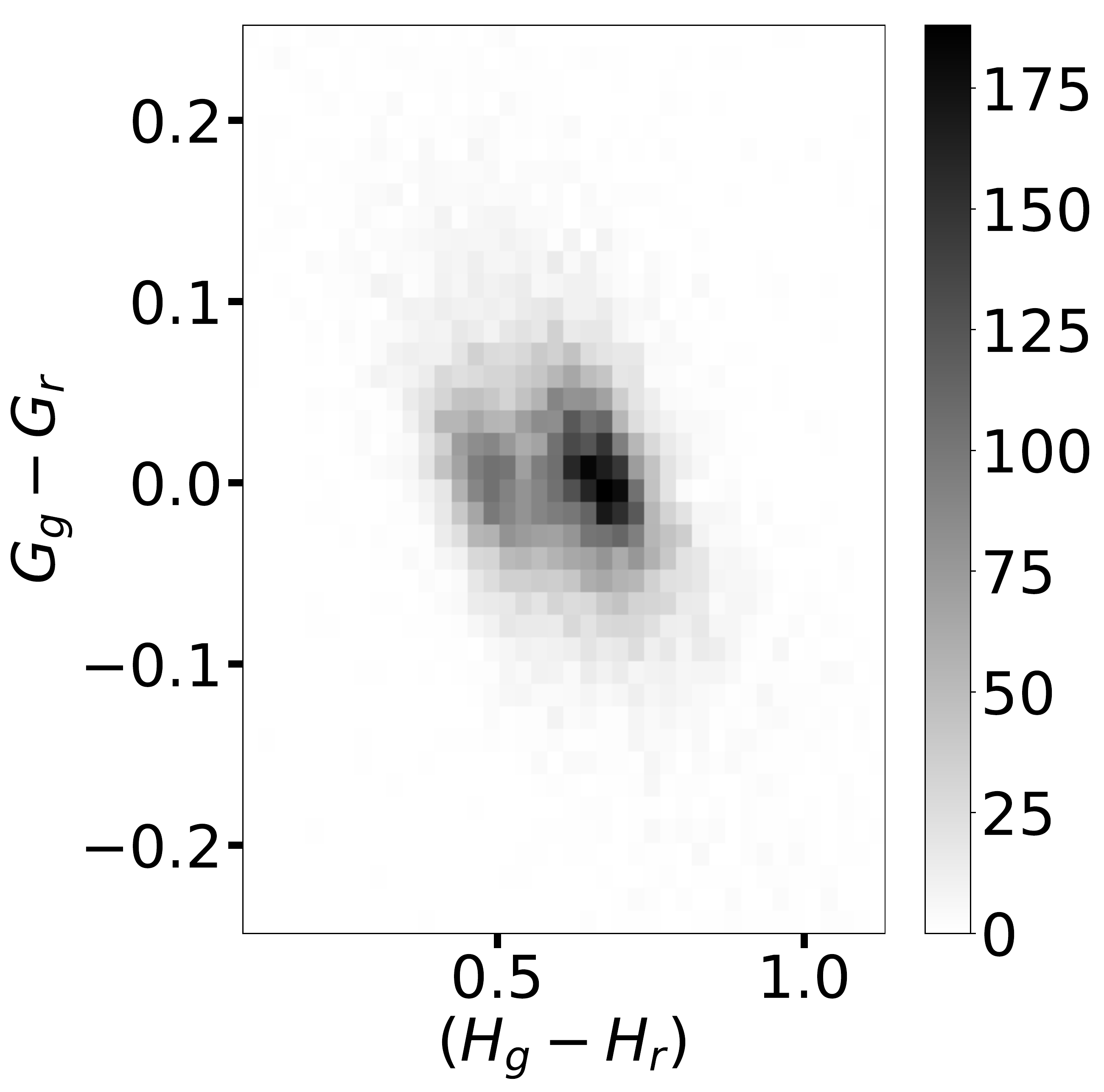}
 \includegraphics[width=6.5cm]{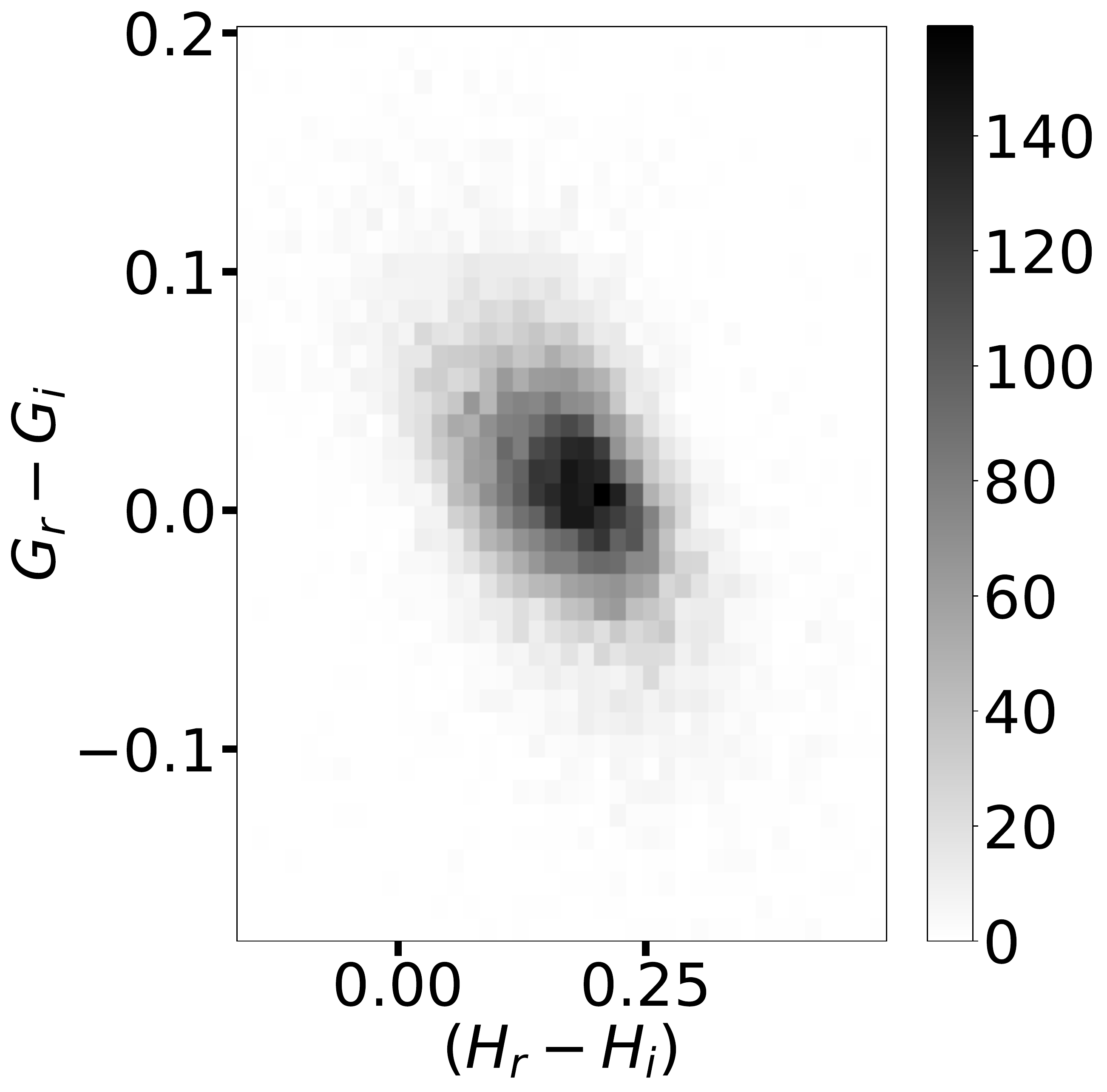}
 \includegraphics[width=6.5cm]{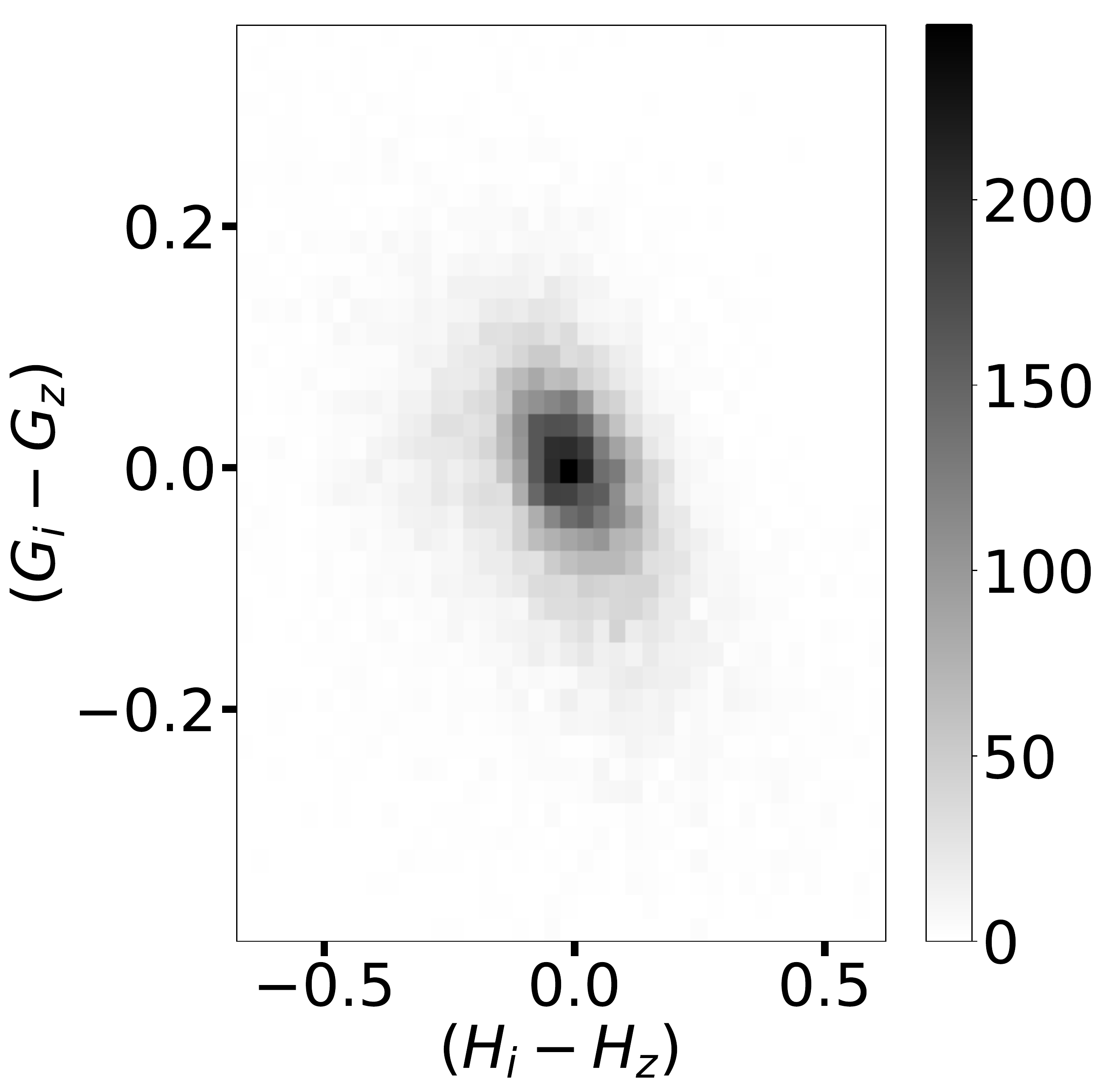}

\caption{Absolute colors vs. the difference $G$ for pairs of filters: Top left, $u-g$; top right, $g-r$; bottom left, $r-i$; bottom right, $i-z$.}\label{fig:H_minus_G}%
\end{figure}

We looked at the possible relationship between the absolute colors and the $G$ values we obtained. In Fig. \ref{fig:H_minus_G} we show the absolute colors vs. the difference in $G$. There seem to be weak anticorrelations (see Table \ref{table:5}) between these quantities, possibly analogous to these found for TNOs \citep{ayala2018,alcan2019}. In the latter work, we proposed it was related to a predominance of single-scattering at a low-phase angle, but the same cannot be said here because, first, $G$ is not the same as the phase coefficient $\beta$ obtained from a simple linear model, and second, the PCs in this work are mostly of asteroids obtained at a wide range of phase angles (see the right panel of Fig. \ref{fig:fig3}).
\begin{table*}
\caption{Results from the Spearman rank-order correlation test.}
\label{table:5}
\centering
\begin{tabular}{c c c}
\hline\hline
Pair & $r_s$ & $ P_{r_s} $ \\
\hline
$(H_u - H_g)$ vs. $(G_u - G_g)$ & -0.27 & 0.0 \\
$(H_g - H_r)$ vs. $(G_g - G_r)$ & -0.38 & 0.0 \\
$(H_r - H_i)$ vs. $(G_r - G_i)$ & -0.45 & 0.0 \\
$(H_i - H_z)$ vs. $(G_i - G_z)$ & -0.37 & 0.0 \\
\hline
\end{tabular}
\end{table*}
Interestingly, in the $g-r$ space, there seem to be two parallel sequences. We do not have a clear explanation yet, but these overall behaviors could be due to some unexplored phenomenon happening at a low-phase angle.

\subsection{Warnings and caveats}\label{sect:caveats}
\begin{enumerate}
\item We explained in Sect. \ref{sec:analysis} that we have used the HG$_{12}^*$ model presented by \cite{penti2016HG}. The $G_{12}^*$ parameter is not well suited for asteroids classed as E and D-type. Nevertheless, we decided to stick to our method because we do not know {\it a priori} the taxonomical classification of our objects. Besides, the results shown in Figs. \ref{fig:fig6} and \ref{fig:fig11} and Sects. \ref{sec:models} and \ref{sec:OE} make us confident about our overall results. However, a revision will be done once we tackle the taxonomical classification of our objects.

\item The values we defined as representative of our quantities and uncertainty interval work well in most cases. Nevertheless, we identified a few cases where they were not the best representatives. In these cases, probably the mean, or mode, could work better. We stress that the method presented here is fully automatic and, due to the number of objects analyzed, it is impossible to perform checks for all objects. The readers have full access, upon request, to the distributions per object and filter of $H$ and $G$ and can therefore perform their analysis. Please note that there are solutions that do not have physical meaning in these distributions, for instance, values of $G<0$.

\item Our method intrinsically assumes that the variations in the brightness in all filters are due to rotational variation and that $P_A(\Delta m)$ is the same in all filters (i.e., due to form rather than different surface properties). We believe this is a reasonable trade-off between massive computation (and large numbers for statistics) and individual targeted studies when using integrated light from point sources, although detailed space missions observations show heterogeneity \cite[e.g.][]{dellagiustina}.

\item We did not apply any correction due to observations taken in different apparitions (changing aspect angle), which may be reflected in the PCs (see Fig. 10 in \citealt{mahlke2021} for an example). In the cases where this could have a significant effect, we expect it to be embedded in the posterior distributions of $H$ and $G$. We stress that we did not search for cases like this.

\item We still do not know the possible rotational phase at which each datum was obtained. This effect may be taken into account by multiplying each value of $\Delta m$ in $P^i(\Delta m|P_j(H_{V})$ by a random $\cos{\phi}$, where $\phi$ is the rotational phase that takes values in $[0,2\pi)$. Preliminary tests show that this approach needs better coverage of the PCs than that provided by the MOC to improve the results.

\item Figure \ref{fig:GAtlas} shows that our values of $G$ are distributed following bell-shape, centered more or less at values associated to C-type asteroids, see Fig. 2 in \cite{penti2016HG}, while it is expected that there should be many values at locations associated to S-type asteroids ($G\lesssim0.2$). This is because our $G$ values are taken as the median of the posterior distributions determined in step 3 in Sect. 3.1.3. The median value is good for statistical analysis of large samples, but it may not be the best option when studying individual objects, as some distributions may appear flat, or bi- (or even multi-) modal.  

\end{enumerate}

\section{Conclusions}\label{sect:conclussions}
We developed a method that includes Monte Carlo simulations and Bayesian inference to include the nominal errors in the magnitudes and the effects of rotational variations in the computation of phase curves with sparse data obtained by large photometric surveys.  We applied the method to the SLOAN MOC and computed the phase curves of almost 15\,000 objects obtaining $H$ and $G$ in the five filter system of the SLOAN survey. We also converted from $g$ and $r$ to $V$ to obtain $H_V$ and $G_V$.

We compared our results, especially $H_V$ and our uncertainties, with published catalogs to evaluate the performance of our method with respect to others. These tests proved satisfactory, granting confidence to the process. Although the method presented was developed with J-PAS and J-PLUS surveys in mind, it is flexible enough to be used with input magnitudes from other surveys, for instance, with data from the LSST, which will observe in a similar photometric system as the MOC and will add a filter in the near-infrared, the y filter  \citep{ivezic2019ApJ}. Therefore, we expect to apply the method to other catalogs, either multi-wavelength or not. Its effectiveness in near-infrared magnitudes, or other wavelengths, has yet to be checked.

Useful information can be extracted from the absolute magnitudes when translating them to reflectance values to construct photo-spectra, while the values of $G$ must be interpreted with care and should not be directly associated with taxonomical information. Finally, we would like to remark that the posterior distributions in $H$ and $G$ can be used when future data comes and other surveys for the same objects as priors to refine the computations and improve uncertainties.

\subsection*{Future work}
The next steps, in no particular order, of this work, include:
\begin{itemize}
\item
cross match with taxonomic information using the absolute colors.
\item
use the zero-phase colors to study color variations with $\alpha$.
\item 
improve our algorithm, including possible effects of changing aspect angle using the shape models derived from Gaia DR3 photometry using inversion methods \citep{cellino2019}. 
\end{itemize}

\begin{acknowledgements}
We thank the input provided by the referee who improved this manuscript. We thank M. Mahlke for useful discussions.

AAC acknowledges support from the Universidad de Alicante under contract UATALENTO18-02, and from the State Agency for Research of the Spanish MCIU through the ``Center of Excellence Severo Ochoa'' award to the Instituto de Astrofísica de Andalucía (SEV-2017-0709).
TSR acknowledges support through grant APOSTD/2019/046 by Generalitat Valenciana (Spain).

Funding for the creation and distribution of the SDSS Archive has been provided by the Alfred P. Sloan Foundation, the Participating Institutions, the National Aeronautics and Space Administration, the National Science Foundation, the U.S. Department of Energy, the Japanese Monbukagakusho, and the Max Planck Society. The SDSS Web site is http://www.sdss.org/.

The SDSS is managed by the Astrophysical Research Consortium (ARC) for the Participating Institutions. The Participating Institutions are The University of Chicago, Fermilab, the Institute for Advanced Study, the Japan Participation Group, The Johns Hopkins University, the Korean Scientist Group, Los Alamos National Laboratory, the Max-Planck-Institute for Astronomy (MPIA), the Max-Planck-Institute for Astrophysics (MPA), New Mexico State University, University of Pittsburgh, University of Portsmouth, Princeton University, the United States Naval Observatory, and the University of Washington.

This work is partially based on data from the SVO MOC Data Access Service at CAB (CSIC-INTA). This work used https://www.python.org/,
https://www.scipy.org/, and
Matplotlib \citep{hunte2007}. 

\end{acknowledgements}

%
%
 \bibliographystyle{aa}
 \bibliography{paper}


\end{document}